\documentclass[twocolumn,useAMS,usenatbib]{aastex62}
\usepackage{graphicx, xcolor}
\usepackage[caption=false]{subfig}
\usepackage{amsmath}

\newcommand{\mstar}{M_{\rm{star}}}
\newcommand{\zstar}{Z_{\rm{star}}}
\newcommand{\SFR}{SFR}

\newcommand{\new}[1]{{{#1}}}

\begin{document}

\title{Comparison of Observed Galaxy Properties with Semianalytic Model Predictions using Machine Learning}

\author{Melanie Simet}
\affiliation{University of California Riverside, 900 University Ave, Riverside, CA 92521, USA}
\affiliation{Jet Propulsion Laboratory, California Institute of Technology, Pasadena, CA 91109, USA}

\author[0000-0003-3691-937X]{Nima Chartab}
\affiliation{University of California Riverside, 900 University Ave, Riverside, CA 92521, USA}

\author[0000-0003-2691-1622]{Yu Lu}
\affiliation{The Observatories, The Carnegie Institution for Science, 813 Santa Barbara Street, Pasadena, CA 91101, USA}

\author{Bahram Mobasher}
\affiliation{University of California Riverside, 900 University Ave, Riverside, CA 92521, USA}

\correspondingauthor{Melanie Simet}
\email{melanie.simet@ucr.edu}

\begin{abstract}

With \new{current and upcoming experiments such as WFIRST, Euclid and LSST, we can observe up to billions of galaxies.}  While such surveys cannot obtain spectra for all observed galaxies, \new{they produce} galaxy magnitudes in color filters. This data set behaves like a high-dimensional nonlinear surface, an excellent target for machine learning.  In this work, we use a lightcone of semianalytic galaxies tuned to match CANDELS observations from \citet{2014ApJ...795..123L} to train a set of neural networks on a set of galaxy physical properties. \new{We add realistic photometric noise} and use trained neural networks to predict stellar masses and average star formation rates on real CANDELS galaxies, comparing our \new{predictions} to SED fitting \new{results}. On semianalytic galaxies, we are nearly competitive with template-fitting methods, with biases of $0.01$ dex for stellar mass, $0.09$ dex for star formation rate, and $0.04$ dex for metallicity.  For the observed CANDELS data, our results are \new{consistent with template fits on the same data at $0.15$ dex bias in $\mstar$ and $0.61$ dex bias in star formation rate}. Some of the bias is driven by SED-fitting limitations, rather than limitations on the training set, \new{and some is intrinsic to the neural network method}. Further errors are likely caused by differences in noise properties between the semianalytic catalogs and data. Our results show that galaxy physical properties can in principle be measured with neural networks at a competitive degree of accuracy and precision to template-fitting methods.

\end{abstract}

\section{Introduction}

Over the last decade, a number of wide-area spectroscopic surveys have been performed, including WiggleZ \citep{2010MNRAS.401.1429D}, BOSS \citep{2013AJ....145...10D}, GAMA \citep{2012MNRAS.427.3244D} and zCOSMOS \citep{2007ApJS..172...70L}. These were used to measure spectral line diagnostics of physical conditions in galaxies to study the evolution of galaxies with look-back time. The results from these studies were often limited by Poisson noise due to the small number of galaxies when divided in terms of their redshifts or in physical parameter bins. In recent years the scale of such surveys have increased by many orders of magnitude in terms of both depth and area coverage. The on-going Dark Energy Survey \citep[DES;][]{2018ApJS..235...33D} and future Large Synoptic Survey Telescope \citep[LSST;][]{2009arXiv0912.0201L} will generate billions of galaxies with multi-waveband photometric data. These will soon join by space-based surveys like Euclid \citep{2011arXiv1110.3193L} and WFIRST  \citep{2015arXiv150303757S}. Given their size, it is impossible to perform spectroscopic observations of individual galaxies. Therefore, techniques should be developed to measure the physical parameters associated with galaxies in these surveys, needed to study the programs they have been designed for. 

The physical parameters for galaxies (redshift, stellar mass, star formation rate, extinction) are conventionally estimated by fitting their observed Spectral Energy Distributions (SEDs) to template SEDs that correspond to the type and class of galaxies in the observed sample. By shifting the model SEDs in redshift space, they are fitted to the observed SED. The physical parameters are then assigned based on the fits, for example by choosing the best-fit template or taking statistics of a probability distribution function of parameters \citep[e.g.]{2000A&A...363..476B, 2008ApJ...686.1503B, Ilbert}. A limitation in using template-fitting methods is that the results depend on a complex array of factors including the type of the galaxies used in template models, the tested parameter ranges, and the filters used in fitting. Also, the estimated physical parameters of galaxies depend on the choice of these initial parameters and the range they cover. For example, if the library of galaxy spectra does not adequately sample the the range of spectra that galaxies can take, there will be serious uncertainties in the estimated parameters for that galaxy; degeneracies between model parameters may produce a good fit with substantial uncertainty in a parameter of interest. Furthermore, apart from redshifts, for which a true estimate (in the form of spectroscopic redshift) could be made for a subsample of objects and used to estimate the uncertainty for the whole population, measuring a “true” value for other physical parameters is difficult and the results often do not directly correspond to the predicted value. Moreover, the results would be affected by the photometric uncertainties, absence of a full photometric coverage and degeneracies between different parameters \citep{2010A&A...523A..31H, 2011MNRAS.417.1891A, 2014MNRAS.445.1482S}. (While this also applies to all methods, including our neural network method described here, the nonlinearity of a neural network changes the dependence on these effects, so depending on the exact degeneracies and uncertainties present, a neural network may have different performance.) Recent studies, such as \citet{2016MNRAS.462.2046G} and \citet{2019MNRAS.486.5104L}, have used mock catalogues generated from galaxy simulations to provide a test data set with known values for many galaxy properties, an approach we adapt here.

Recently, Machine Learning (ML) techniques have emerged as independent alternatives for measuring the physical parameters of galaxies \citep{2016PASP..128j4502S, 2008ApJ...683...12B, 2015MNRAS.449.2040H, 2015ApJ...813...53M}. Using a training sample with known physical parameters, they generate statistical models to predict the distribution of those parameters in a target data set. The ML techniques are only applicable within the limits of the training data. Any extrapolation of them to a different redshift or mass regime would lead to errors in the final estimate. However, a distinct advantage of ML is that one could incorporate extra information (i.e. morphology, galaxy light profile) within the algorithm.  The ML techniques are divided into two categories: supervised and unsupervised.  In supervised ML, the input attributes (magnitudes and colors) are provided along with the output (redshift) and directly used for training in the learning process \citep{2008MNRAS.390..118L, 2009MNRAS.398.2012F, 2010ApJ...715..823G}. Here, the learning process is supervised by the input parameters. The unsupervised techniques do not use the desired output values (e.g. spectroscopic redshifts) for training purposes, and are less frequently used.  

The use of ML methods to study galaxies in survey data has so far been focused on measuring photometric redshifts of galaxiesm with a few studies addressing other properties. This is because the ML algorithms can be more easily trained using the spectroscopic redshifts whereas for other parameters such “true” and unambiguous estimates are not possible to obtain without making serious assumptions. Detailed comparisons have been performed between different methods and algorithms measuring photometric redshifts \citep{2013MNRAS.432.1483C, 2013ApJ...775...93D, 2017MNRAS.468.4323B} and masses \citep{2015ApJ...808..101M} of galaxies. However, except in a few cases where ML techniques are used (MCC and neural net) these were mostly based on different variants of template fitting using observed or model SEDs.  ML methods have also been used to study a diverse portfolio of galaxy properties besides redshift, including morphology \citep[e.g.][]{2018MNRAS.477..894A, 2018arXiv180710406D, 2019arXiv190107047B, 2019arXiv191106259C, 2019PhLB..795..248K, 2020MNRAS.491.1554W}, star formation rate \citep[e.g.][]{2017MNRAS.464.2577S, 2019A&A...622A.137B, 2019MNRAS.486.1377D}, and stellar masses \citep[e.g.]{2019A&A...622A.137B, 2019A&A...624A.102D}  as well as to detect unusual objects \citep[e.g.][]{2019BSRSL..88..174S}. Alternate data vectors have also been used, such as in \citet{2019A&A...621A..26P} who use the per-filter pixel counts in multicolor postage stamps as their numerical data rather than measurements of total flux. Several approaches use unsupervised ML techniques to first compress the high-dimensional color-space information (through projection or selection) and then use functions of the compressed data vector to predict one or more physical properties \citep{2017MNRAS.468.4323B, 2017MNRAS.464.2577S, 2019ApJ...881L..14H, 2019MNRAS.489.4817D, 2019arXiv190909632W}. ML methods can also be used to improve estimates from other techniques. For example,  \citet{2014MNRAS.442.3380C} used a Bayesian combination of photometric redshift Probability Density Functions (PDFs) using different ML methods to improve estimates of photometric redshifts.  ML techniques are also attractive because they are computationally more efficient: once a method has been trained, it can be deployed cheaply on many galaxies (with computations typically being linear or polynomial), resulting in a much shorter computation time per galaxy than SED-fitting.

In this paper we use a machine learning technique, a neural network, to measure the stellar mass and star formation rates of galaxies with available photometric data and photometric redshifts using a training set of semianalytic galaxies from \citep{2014ApJ...795..123L}. The test sample in this study is from the Hubble Space Telescope (HST) Cosmic Assembly Near-infrared Deep Legacy Survey (CANDELS) GOODS-South field. This has imaging data in optical (ACS) and near-infrared (WFC3) wavelengths as well as Spitzer Space Telescope mid-infrared (IRAC) bands. We describe the CANDELS data and semianalytic galaxies in section~\ref{sec:data}, give an overview of the neural network ML algorithm we use in section~\ref{sec:ml}, test our neural networks with a semianalytic test set in section~\ref{sec:nnperf}, \new{compare neural networks and template fitting on a mock data set in section~\ref{sec:nn_sed_comp}}, and finally apply our networks to CANDELS data in section~\ref{sec:mldata}, with summary and conclusions in section~\ref{sec:summary}.  Throughout the paper, we assume the cosmology of the simulation used to generate the mock catalog and semianalytic galaxies: $\Omega_M=0.27$, $\Omega_{\Lambda}=0.73$, $\Omega_b=0.044$, $h=0.7$, $n=0.95$, and $\sigma_8=0.82$.  

\section{Data}\label{sec:data}

\subsection{CANDELS galaxies}\label{sec:data:candels}

The Cosmic Assembly Near-IR Deep Extragalactic Legacy Survey \citep[CANDELS;][]{2011ApJS..197...35G,2011ApJS..197...36K} is a $\sim 800$ arcmin$^2$ survey performed using the Wide Field Camera 3 (WFC3) and Advanced Camera for Surveys (ACS) on the Hubble Space Telescope.  The survey consists of five fields (GOODS-S, GOODS-N, COSMOS, EGS, and UDS). Multi-waveband photometric imaging observations were performed spanning the wavelength range from UV to mid-infrared. In two of these fields (GOODS-S and GOODS-N) deeper photometric observations over smaller areas were performed.

In this study, we use data from the GOODS-S deep field \citep{2013ApJS..207...24G} covering an area of 170 arcmin$^2$. The combination of CANDELS pointings with supplementary data sets \citep{2004ApJ...600L..93G,2007ApJ...659...98R,2010ApJ...709L.133B,2011ApJS..193...27W} results in a multi-waveband catalog consisting of three WFC3 filters (F105W, F125W and F160W) and five ACS filters (F435W, F606W, F775W, F814W and F850LP). Any galaxy that was not detected in any of the above filters or which was not covered by imaging in any of the filters was excluded from the catalog.  The final data set consists of 20,512 objects out of an initial 34,930. 

We use the spectral energy distribution (SED) fitting code {\sc LePhare} \citep{Arnouts,Ilbert}, combined with the \citet{BC03} stellar population synthesis models, to derive the physical properties for each galaxy (stellar mass and star formation rate). We assume an exponentially declining star formation history with nine different e-folding times in the range of $0.01<\tau<30\ \rm{Gyr}$. The dust properties are modeled with varying $E(B-V)$ between 0 and 1.1 assuming the \citet{Calzetti} dust attentuation curve. We also take into account nebular emission line contribution as described in \citet{Ilbert2009} and assume a Chabrier initial mass function \citep{Chabrier}, with lower and upper mass limits 0.01 Msun and 100 Msun respectively.  We consider three different metallicity values: $Z = 0.02$, $0.008$, and $0.004$.

{\sc LePhare} produces a marginalized likelihood of stellar masses, and we use the median value of this likelihood as our stellar mass estimate \citep{chartab20}.  To measure the star formation rate (SFR), we use the rest-frame UV luminosity which traces timescales of $\sim 100$ Myr which is associated with continuum from massive, short-lived $\rm O$ and $\rm B$ type stars. We adopt the \citet{Salim} SFR(UV) calibration: 
\begin{equation}\label{eq:SFR}
\log_{10} \rm SFR=-0.4 M_{UV,AB}-7.53
\end{equation}
where M$_{\rm{UV,AB}}$ is the dust-corrected monochromatic absolute UV magnitude in AB system. We measure the observed UV magnitude ($\rm M_{UV,observed}$) by using the 1600$\mbox{\AA}$~flux density from the best-fit SED. The UV spectral slope ($\beta_{\rm UV}$) is measured by fitting a power law of the form $f_\lambda \propto \lambda^{\beta_{\rm{UV}}}$ between $1300{\rm \mbox{\AA}} < \lambda < 2600 {\rm \mbox{\AA}}$ to the best-fit SED, with $f_\lambda$ being the wavelength-dependent flux density. We dust correct the observed M$_{\rm{UV}}$ by assuming the \citet{Meurer} calibration:
\begin{equation}\label{eq:SFR}
\rm M_{UV}=\rm M_{UV,observed}-(1.99\beta_{\rm UV}+4.43)
\end{equation}
where M$_{\rm{UV}}$ is the dust-corrected UV magnitude. We use the \citealt{Meurer} calibration to find the dust-corrected UV since it only requires a UV measurement, rather than the \cite{Calzetti} attenuation curve which needs E(B-V) measurements and is limited by the resolution of E(B-V) parameter being searched in SED fitting.

\subsection{Semianalytic galaxies}\label{sec:data:semianalytic}

We use mock catalogs generated from semianalytic models to mimic CANDELS observations, as presented in \citet{2014ApJ...795..123L}.  In short, a semianalytic model takes a cosmological dark-matter-only simulation and adds baryonic components with recipes for their evolution through cosmic time depending on the evolving properties of the dark matter host halo.  This baryonic component can consist of stars, cold gas, and hot gas with various physical processes transferring mass from one component into another (e.g., stars can form from cold gas).    

The mock catalog we use was presented in \citet{2014ApJ...795..123L} as the ``Lu'' model. It assumes heating of the gas by reionization which, in turn, limits the fraction of the baryons that collapse into the halos \citep{2000ApJ...542..535G, 2004ApJ...609..482K}. Radiative cooling is estimated assuming the \citet{2006MNRAS.365...11C} model that collapses a fraction of the hot gas onto central (but not satellite) galaxies depending on the cooling timescale of the halo.  As in other semianalytic models, the cold gas is assumed to be distributed in an exponential disk where stars are formed.  A particular feature of model we use here is that the star formation rate depends on the circular velocity of the host halo in addition to the more typical dependences on star-forming gas mass and disk dynamical time scale.  Supernova feedback reheats the cold gas and ejects both cold and hot gas.  No explicit model for black hole accretion or AGN feedback is assumed, but a halo quenching model is implemented that switches off radiative cooling above a halo mass around $10^{12} M_{\odot}$.  Galaxy mergers are handled by following subhalo information in the merger tree and assuming that even an unresolved subhalo will remain intact for some fraction of the dynamical friction time.  

A fraction of the mass in new stars is assumed to convert into metals and is instantly recycled back into the disk (and from there into the cold and hot gas surrounding the disk, according to the above prescriptions), parameterized using a Chabrier initial mass function for stellar mass in the range of $0.1 M_{\odot}$ and $100M_{\odot}$ ~\citep{2003PASP..115..763C}. The model therefore explicitly predicts the star formation history and the metallicity enrichment history of a model galaxy. These histories are converted into the intrinsic spectrum energy distribution (SED) for each mock galaxy by adopting the CB07 stellar population synthesis model ~\citet{CB07}, which is an updated version of the BC03 model ~\citep{BC03}. The observed magnitudes are computed by integrating the SED that is redshifted according the redshift of the galaxy and weighted by the transmission functions of observational filters. We also adopt the \citet{Charlot2000} model to account for dust attenuation and predict dust attenuated colors.

The merger trees for our mock catalogs used were drawn from the Bolshoi $N$-body cosmological simulation~\citep{2011ApJ...740..102K}, with a volume of (250 Mpc/$h$)$^3$, using 8 billion particles with a mass resolution of $1.35\times 10^8 M_{\odot}$, and 180 stored time steps.  Halo finding was performed with the Rockstar code~\citep{2013ApJ...762..109B} and merger trees were constructed using the Consistent Trees algorithm~\citep{2013ApJ...763...18B}. 
Lightcone halo mock catalogs are extracted from the simulation box. These lightcone catalogs mimic the five CANDELS fields, and have redshift range from $z=0$ to $z=10$. Eight realizations are generated for each of the fields. Each dark matter halo in a lightcone catalog has a unique ID. The corresponding dark matter halo merger tree branch is found from the simulation box and rooted on the halo.

The model parameters used in \citet{2014ApJ...795..123L} are tuned to match calibration data.  The main calibration set is the stellar mass function of local galaxies from~\citet{2013ApJ...767...50M}.  Parameters were tuned using Markov Chain Monte Carlo chains with the differential evolution algorithm~\citep{Braak2006} and a likelihood based on a weighted $\chi^2$, with a parameterization to account for incompleteness in the data at low mass.

The best fit model is adopted to apply onto each merger tree of the lightcone mocks to predict the star formation history of each galaxy hosted by every halo in the mock. The semianalytic models contain simulated magnitudes for the eight CANDELS bands described above.

We only use galaxies from the mock catalog within the redshift range $0.1 < z < 6$. We also impose a cut such that $H_{AB}<30$ mag, to avoid a small population of faint high-mass galaxies that are separated in parameter space from the rest of the models, and are significantly fainter than them.

In the real world, the data contains observational uncertainties. To examine how much observational error degrades the best-case performance of our neural networks, we will also analyze the simulations with pseudo-``observational error'' applied. We incorporate errors in the mock catalogs using the observational errors associated with CANDELS galaxies. We measure the median multiplicative flux error in bins of magnitude with width 0.5. These were then linearly interpolated to obtain a typical multiplicative uncertainty for a given magnitude in the data. We then draw a random Gaussian-distributed number $\delta F$ with a scale length given by this multiplicative uncertainty, and add the term $\eta = -2.5\log_{10} (1+\delta F)$ to simulate observational errors.  The simulation data perturbed in this way has distributions of magnitudes and colors similar to the CANDELS data.  This pseudo-observational error is added when the catalogs are read into our neural network software, before any colors are computed or any splitting into training and validation data sets is performed.  We also explore a case where we mimic the noise properties of only well-resolved galaxies, defined as galaxies with flux signal-to-noise ratios $\geq 5$ in each observed band.  In both cases, after we have perturbed the magnitudes, we apply a cut to remove galaxies that have scattered below the minimum observed magnitude in each band.

\subsection{Comparison of semianalytic models with CANDELS observations}

The semianalytic models have been tuned to match observations \citep{2014ApJ...795..123L}.  Here, we compare the galaxy magnitudes in both the noise-free and noisy case, to demonstrate the feasibility of using the semianalytic galaxies to train a neural network that is then applied to the CANDELS data.

\begin{figure*}
\begin{tabular}{cc}
    \centering
    (a) & (b) \\
	\includegraphics[width=0.45\textwidth]{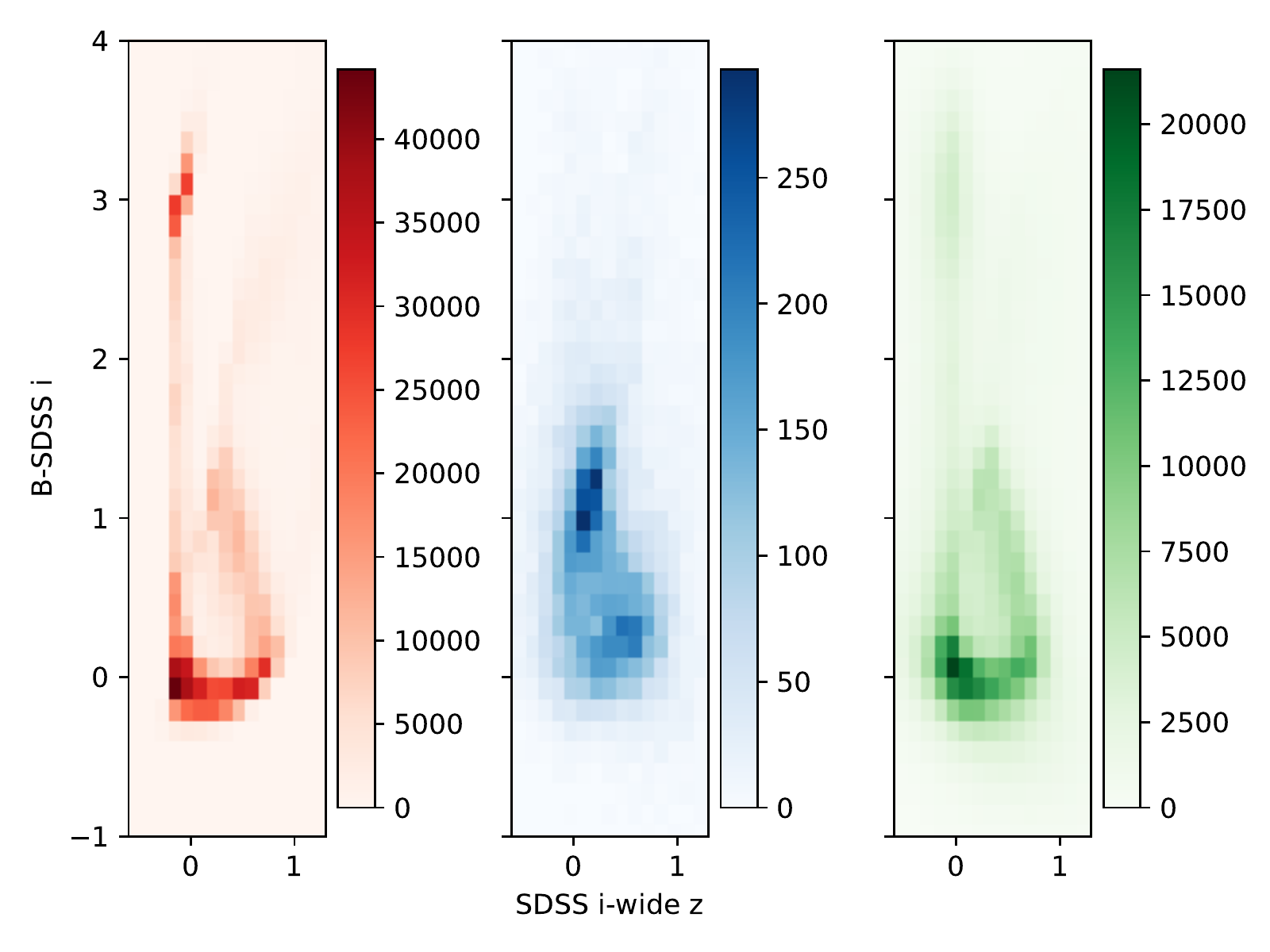} & 
		\includegraphics[width=0.45\textwidth]{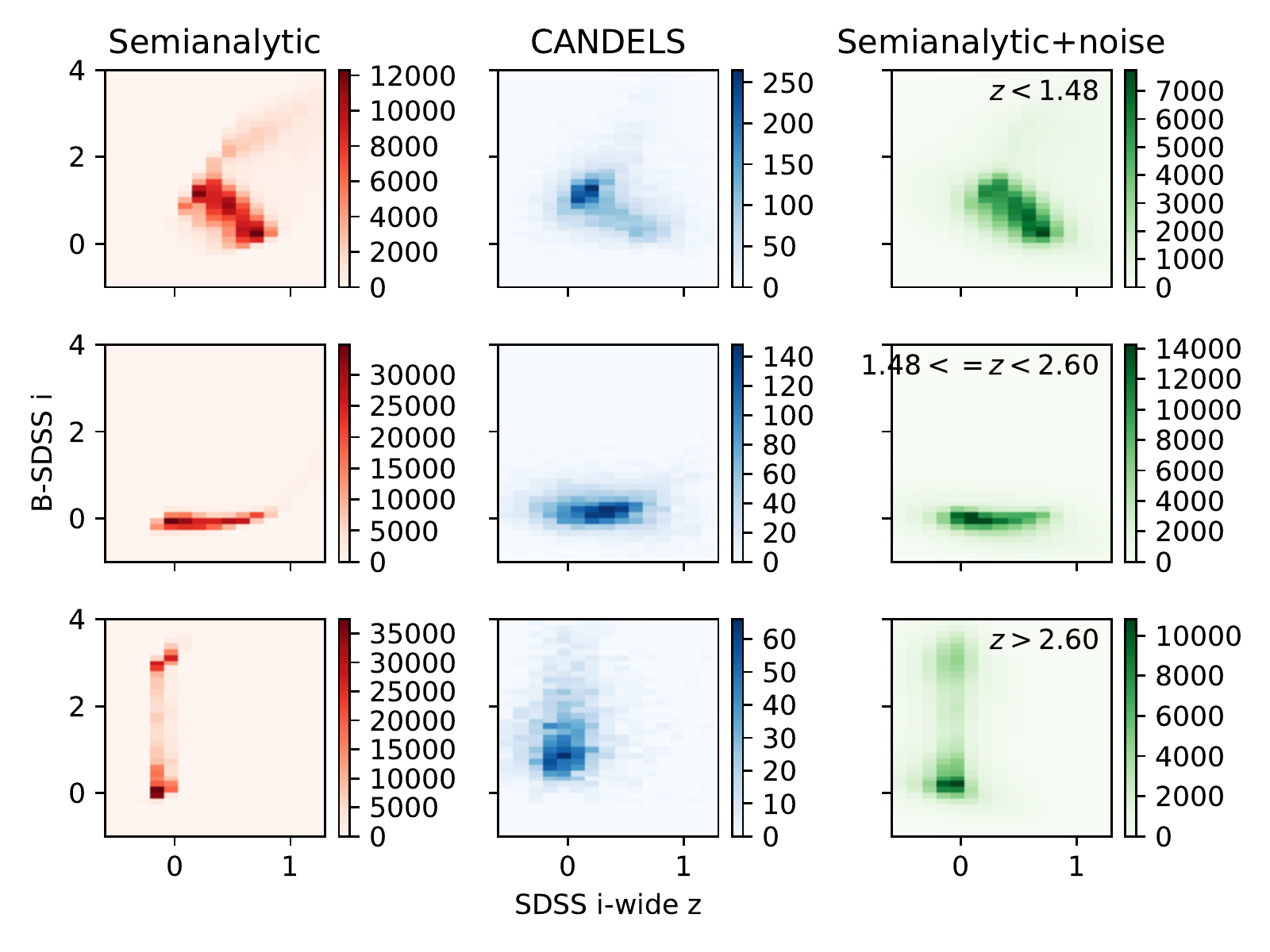}
	\end{tabular}
	\caption{(a) Color-color plots for one projection of the 8-dimensional color space. Left panel contains semianalytic galaxies without noise, center panel is CANDELS measurements, right panel is semianalytic galaxies with simulated photometric noise.  The spanned color space is very similar; the vertical feature in the semianalytic plots missing from the CANDELS distribution consists of faint high-redshift galaxies below the CANDELS completeness limit.  The difference in color-space occupation also reflects incompleteness in the CANDELS data.  The semianalytic galaxy catalogue and the CANDELS catalogue are both cut at a redshift $z=6$ (in the CANDELS case, from our template-fitting procedure), but the incompleteness at faint magnitudes means the CANDELS redshifts have relatively more low-redshift galaxies and relatively fewer high-redshift galaxies.  (b) As panel (a), but the three rows represent a split into three equal-population bins of redshift in the semianalytic catalogue. Here it is more clear that the galaxy manifolds are in the same location (accounting for noise) and the differences in panel (a) are differences in redshift distributions.}\label{fig:cc}
\end{figure*}

\begin{figure}
    \centering
	\includegraphics[width=0.45\textwidth]{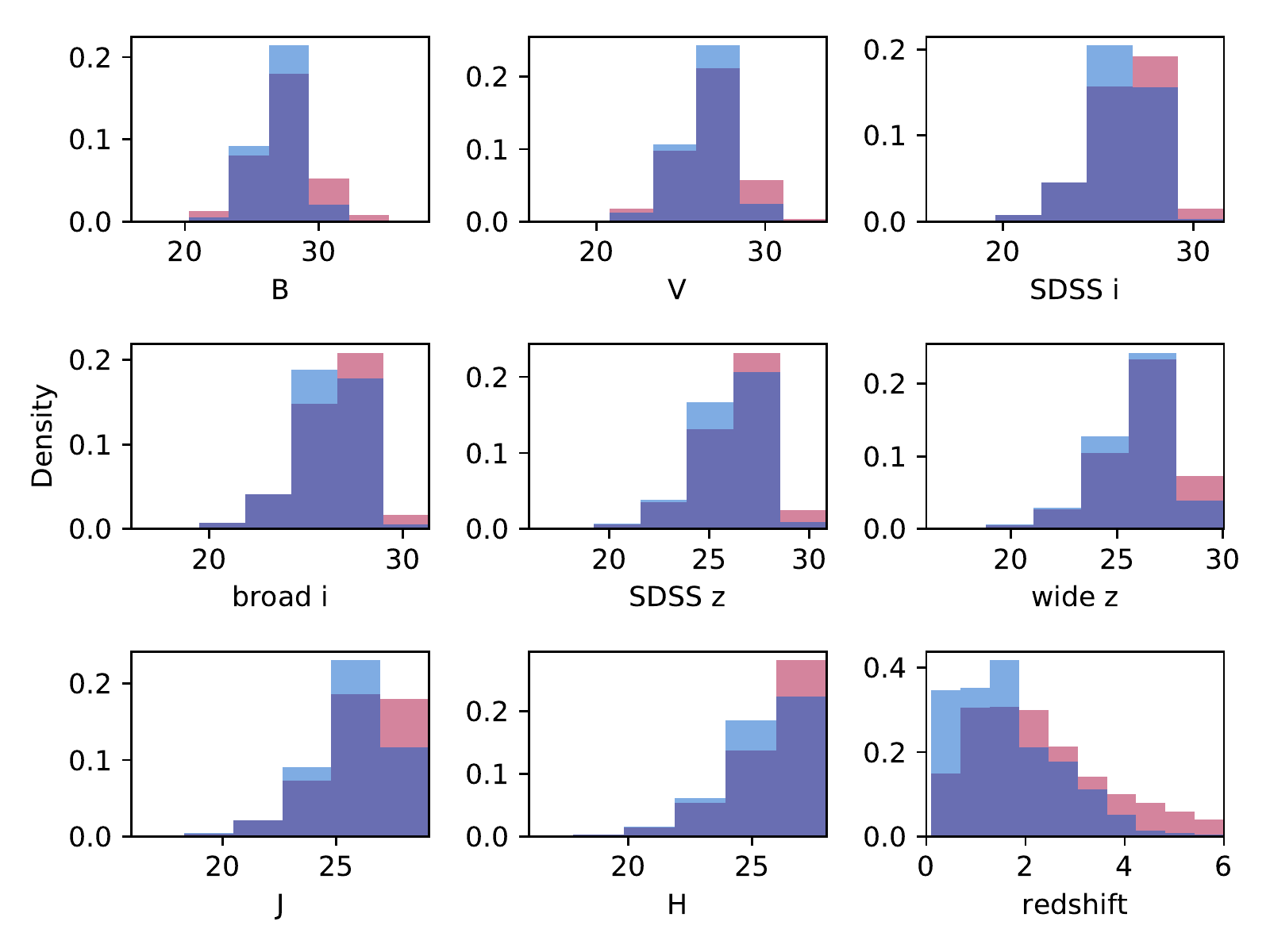}
\caption{Histograms of the distribution of semianalytic and CANDELS galaxies in our eight bands and in redshift.  As in Fig.~\ref{fig:cc}, red represents the semianalytic galaxies and blue represents the CANDELS data. The spanned ranges are similar, but the semianalytic galaxies in all cases have better completeness at fainter magnitudes and higher redshifts, due to the lack of observational selection functions.}\label{fig:cc}
\end{figure}

In Fig.~\ref{fig:cc}, we show color-color plots, with heatmaps showing the location of both the semianalytic galaxies (with and without simulated noise) and the observed CANDELS galaxies. We chose this projection of the higher-dimensional color space as good representatives of the general trend: the semianalytic galaxies without noise lie on thin manifolds, with the CANDELS galaxies being consistent with their overall trend but significantly broadened by observational noise, and the semianalytic galaxies with added noise looking similar to the CANDELS galaxies, but with some galaxies smeared outside the boundaries of the semianalytic and CANDELS galaxies by the noise.  In the first panel, the vertical manifold at $i-z \approx -0.2$ in the semianalytic galaxies does not show up strongly in the data. However, this is expected, as the galaxies on that part of the manifold are at higher redshift and fainter, below the CANDELS completeness limit.

We do not expect our machine learning method to be affected by the presence of objects in our semianalytic sample that do not appear in the CANDELS data, given that such objects are less frequent in number than representative objects. It is more important that all CANDELS galaxies have good representation in the semianalytic data set.  Having semianalytic objects outside the bounds of the CANDELS galaxies merely means that we have trained a machine learning method for data it will never see.  The only exception would be if the faint semianalytic galaxies introduced a degeneracy in the color-space manifold that is not there for brighter galaxies.  However, we do not see signs of this in our data set (the quality of the fit to brighter galaxies does not improve when excluding fainter galaxies).
have 
\section{Machine learning procedure}\label{sec:ml}

Machine learning techniques have recently been adapted to problems in  astronomy including photometric redshifts \citep[e.g.][]{2014MNRAS.445.1482S, 2017arXiv170904205B, 2018PASJ...70S...9T}, large-scale structure \citep{2018arXiv180400816A}, galaxy morphologies \citep{2018MNRAS.476.3661D}, and calibration factors for weak lensing data measurement algorithms~\citep{2018arXiv180702120T}. In this work, we use the technique known as a neural network.

\begin{figure*}
	\includegraphics[width=0.95\textwidth, clip=True, trim=5cm 5cm 4cm 5cm]{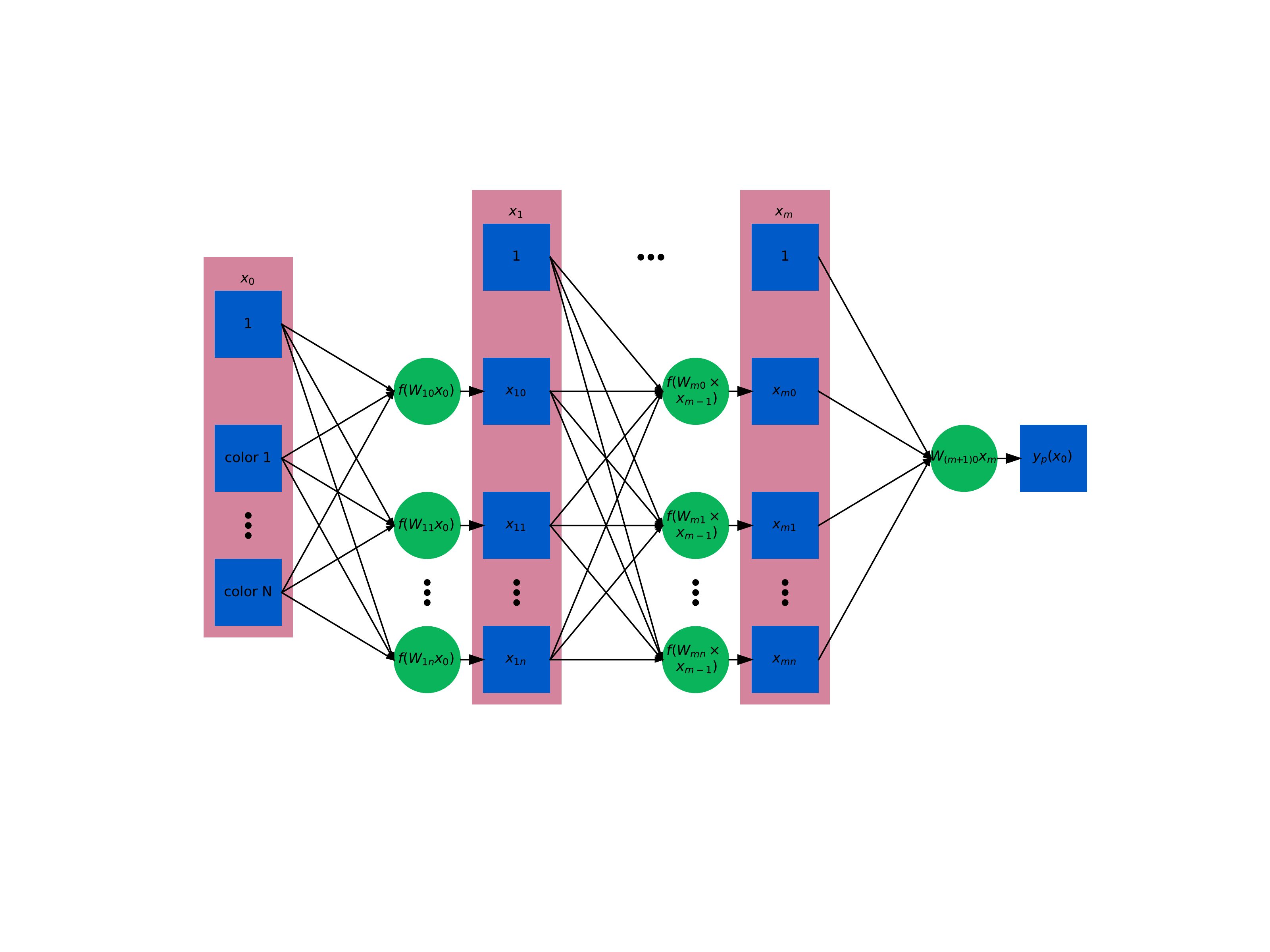}
	\caption{A schematic for a simple neural network.  On the left, the inputs consist of $N$ galaxy colors shown in blue boxes, plus a constant value to provide offsets.  This vector of data (everything in the red rectangle labeled $x_0$) is passed to a layer of $n$ nodes (the green circles); each node performs a weighted sum, and then performs a nonlinear transformation of the sum via $f(x)$.  The outputs of this layer of nodes, along with another constant value, form a new vector, $x_1$, which is itself fed to another layer of $n$ nodes.  This process repeats until we have fed the data through $m$ layers.  Finally, the outputs of the last layer, $x_m$, are fed to a single node that performs a weighted sum, and the value of this weighted sum is our prediction, $y_p$.}\label{fig:nn_layout}
\end{figure*}

The idea of a neural network is simple.  The underlying motivation is that we have a data vector (say galaxy colors), as well as a desired output (stellar mass, for example) that is a nonlinear function of that data vector, and we want to learn how to estimate the output given a new data vector.  We don't know the optimal form of the function, however, and even if we did, it is likely that the complicated form would make determining the parameter values analytically difficult. Neural networks computationally determine the important combinations of the data points, and the appropriate coefficients for those combinations, by breaking the prediction process down into a series of linear combinations of the data points, combined with nonlinear transformations of those sums to reproduce nonlinear behavior.  

As illustrated in Fig.~\ref{fig:nn_layout} for a single data point, in the usual neural network setup, a set of vectors is fed to a layer of computational units called nodes.  Each node computes a weighted sum of the coordinates of each vector, and optionally applies a simple nonlinear function to the sum (which might clip negative values, for example). The output of the layer--one value for each node for each input datapoint--can then act as a new set of vectors, which is fed to a new layer of nodes.  The final layer has only one node, and the value it computes for each data point is the prediction, $y_p$, that corresponds to the true value of $y$ for the data points.  The free parameters of the model produced by the network are the weights used to compute the weighted sum in each node.  To compute the optimal values of these parameters---a process called \textit{training the network}---iterative adjustments are made based on a comparison of the predictions $y_p$ to the known true values for the input data points $y$.  The iterations proceed until either the desired accuracy is reached or the network stops improving its accuracy.   A more detailed overview of this technique with additional complications can be found in  \citet{Goodfellow-et-al-2016}.  Error estimation can be performed by training multiple networks~\citep[e.g.][]{2017arXiv170904205B} but here we use a simple point estimate from a single network.  
In its simplest form, a neural network consists of a series of matrix multiplications with a simple function applied to the outputs: for an input data point $x$ with $M$ features fed to a layer of $N$ nodes, the output $O_{i}$ of the $i$th layer is simply
\begin{equation}\label{eq:NN}
O_i = f(W_i x + C_i)
\end{equation}
where $C_i$ is a scalar or vector of offsets and $W_i$ is a matrix.  The function $f(x)$ is chosen to fit the problem at hand, while the values of the matrices $W_i$ are numerically trained to fit the training sample as described above. By convention, this function is called a ``response function'' in machine learning terminology. Again, the response function is usually nonlinear: if it is linear, then the whole network produces a simple linear combination of the input data. The computational difficulty of neural networks comes from the complexity of training the many values of the $W$ matrices, and particularly the difficulty of training more than one or two layers \citep[and references therein]{Goodfellow-et-al-2016,2018MNRAS.473.3895L}.  The initial values of $W$ are typically set near 1 with small random offsets in each element \citep{tensorflow2015-whitepaper}, meaning that (unless the random number generator is seeded in the exact same way) the same neural network architecture will produce slightly different predictions for the same training set. 

The necessary choices to set up a neural network include:
\begin{itemize}
	\item The number of layers
	\item The number of nodes in each layer
	\item The response functions $f_i$
	\item The method used to train the network (i.e., to alter the weight matrices $W_i$ after each iteration)
	\item The function that computes a metric distance between the predicted points and the true values (the \textit{loss function})
\end{itemize}

We use the Google package \texttt{TensorFlow}\footnote{\url{https://www.tensorflow.org/}} \citep{tensorflow2015-whitepaper} to implement our neural network. \texttt{TensorFlow} is a highly-optimized framework designed to enable fast implementation of neural networks and other machine-learning problems. The package runs mostly compiled code to increase speed of execution, but the user interface is in Python.  For this work, we use the high-level ``Estimator'' API for \texttt{TensorFlow}, which automates most of the bookkeeping necessary to setup and train a neural network. We began with a set of 2 10-node layers, as used before for photometric redshifts in e.g. \citet{2004PASP..116..345C}, and added nodes and layers until our performance (measured by the value of the loss function) stopped improving.  We found optimal performance with a set of 3 20-node layers; we found this architecture to be complex enough to reproduce high-dimensional nonlinear manifolds in color space, while still being simple enough that the neural network training algorithm could converge on a good solution. The response functions $f_i$ in our network are the ``relu'', or rectified linear unit, function \citep{pmlr-v15-glorot11a}:
\begin{equation}\label{relu}
f(x)=\begin{cases}
x&  \text{if $x>=0$},\\
0&  \text{if $x<0$}.
\end{cases}
\end{equation}
This satisfies the requirement that $f(x)$ is nonlinear in $x$ to reproduce nonlinear behavior, while still being computationally efficient.

Our loss function is a simple squared distance between the labels $y$ from our catalog and the predictions $y_p(x)$ from a given training step of our neural network, summed over input training points $k$:
\begin{equation}\label{eq:lossfunc}
L = \sum_k \left(y_k - y_p(x_k)\right)^2
\end{equation}
This is related to the $\chi^2$ that is more commonly used.  In this case, we do not normalize by the expected values $y_k$ or expected uncertainty; we found that doing so has the effect of focusing the training first on small values of $y_k$, in practice making it more difficult to find a manifold that works well for the entire parameter range.

\subsection{Training the network}

To train the network, we split the data into two sets: a set to use to train the coefficients of the network, and a set to evaluate how well our network is reproducing the data.  The two sets must be different in order to avoid overfitting, the problem where the network reproduces not only the average trends in the data, but the specific noise fluctuations of the data set used for training \citep{Goodfellow-et-al-2016}.  We use a 70-30 split, with 70\% of the data used to train the coefficients (called the ``training set'') and 30\% to check its performance at intervals (called the ``validation set'').  The galaxy color manifold is well-sampled in this simulated data set, such that this simple random split produces nearby neighbors from the training set for all data points in the validation set (relative to the distribution in the full data set), even in relatively sparse regions of color space.

We use stochastic gradient descent to train the network, in which we iteratively compute the predictions of the network on a random subset of our training data \citep{Goodfellow-et-al-2016}.  In our case, we find that using $10,000$ points per step works well, with a check against the validation set every $500$ steps ($5\cdot 10^6$ total training data points passed through the network).  We use the Adam optimizer \citep{DBLP:journals/corr/KingmaB14} to update the coefficients after every step.  Briefly, the Adam algorithm involves an adaptive learning rate computed from moments of derivatives of the loss function, Eq.~\eqref{eq:lossfunc} so that the coefficients change more quickly when the gradient of the loss function is high, and then change only slowly as the loss function approaches a (local) minimum. We use the default parameter settings for the Adam optimizer implementation within \textsc{TensorFlow}. We explored changing the learning rate $\alpha$, which controls how fast or slow the coefficients change for a given value of the loss function, but accuracy and precision decreased as we moved away from the default value.

\section{Application to semianalytic galaxies}\label{sec:nnperf}

In our semianalytic catalog, each mock galaxy is associated with four physical parameters: stellar mass $\mstar$, redshift $z$, stellar metallicity $\zstar$, and average star formation rate $\SFR$.\footnote{Other quantities in the catalog, for example the dark matter halo mass, will correlate with galaxy light properties because they correlate with, for example, the stellar mass; for this work, we consider only the direct correlations, not such indirect correlations, which pick up both additional parameters like the stellar-halo mass connection and additional noise.} There has been good progress in using machine learning techniques to measure photometric redshifts~\citep{2018NatAs.tmp...68S}. Therefore, we concentrate on measuring other parameters here.

We train networks using 5 possible sets of input data columns.  We will use the bold letters in parentheses as shorthand in tables throughout the paper.
\begin{itemize}
	\item Galaxy magnitudes (\textbf{m})
	\item Galaxy pairwise colors ($B-V$, $V-i$, etc) (\textbf{c})
	\item Galaxy magnitudes and pairwise colors (\textbf{mc})
	\item All galaxy colors ($B-i$ in addition to $B-V$ and $V-i$, etc) (\textbf{C})
	\item Galaxy magnitudes and all galaxy colors (\textbf{mC})
\end{itemize}

Machine learning algorithms generally respond to information in different ways than the deterministic model-fitting methods more commonly used in astronomy.  If the output is sensitive to a particular combination of data points (such as a color formed from two non-adjacent filters), then it is often more numerically efficient to give the algorithm this combination, even though the network could eventually learn that the given combination is important. The ML training algorithm may converge on an answer more quickly if relevant data combinations are given as inputs, since it does not have to learn that, for example, $g-r$ is an important combination of filters, before it learns exactly how $g-r$ relates to the physical parameter of interest. To some extent this is also dependent upon the training we are doing: we are, in some sense, optimizing the inputs for the architecture of our network, in addition to optimizing the network that uses those inputs.  A different set of layer and node parameters, or a different learning rate, for example, might respond differently to the choice of input values. 

To quantify how well a networks performs, in addition to the loss function, we will report the mean bias, uncertainty, and 3$\sigma$ outlier rate, computed for the parameter
\begin{equation}
b_k = y_k-y_p(x_k)
\end{equation}
where, as above, $y_k$ is the true value of the galaxy property for data point $k$, and $y_p(x_k)$ is the predicted value for the galaxy property computed using the vector $x_k$.  We report $\langle b \rangle$, $\sigma_b=(\langle b^2\rangle-\langle b \rangle^2)^{1/2}$, and the fraction of data points with $|(b_k-\langle b \rangle)/\sigma_b|>3$.  We select the set of inputs with the minimum loss function as our optimal set, as this value is sensitive to both the bias and the uncertainty.

In this section, we will train neural networks for each of the main galaxy properties, with these 5 different input column choices, and with and without simulated photometric noise. We then analyze the bias and uncertainties in the results.

\subsection{Basic predictions}
\subsubsection{Stellar mass}

\begin{table*}
\begin{center}
	\begin{tabular}{cccccccc}

 Noise & Input & Number &  & Average &  & 3$\sigma$ outlier \\
 level & columns & of steps & Loss$^{1/2}$ & bias & Uncertainty & rate  \\
\tableline
None & m & 100000 & 0.157 & $-0.055$ & 0.147 & 0.013 \\
None & c & 100000 & 0.115 & 0.003 & 0.115 & 0.016 \\
None & C & 100000 & 0.120 & $-0.021$ & 0.119 & 0.018 \\
None & mc & 100000 & 0.071 & $-0.025$ & 0.066 & 0.0147 \\
\textbf{None} & \textbf{mC} & $\mathbf{100000}$ & $\mathbf{0.058}$ & $\mathbf{-0.011}$ & $\mathbf{0.056}$ & $\mathbf{0.015}$\\
\tableline
5$\sigma$ & mC & 10000 & 0.124 & 0.007 & 0.124 & 0.012 \\
Full noise & mC & 10000 & 0.204 & -0.019 & 0.203 &  0.018\\
\end{tabular}
	\caption{Summary of the performance of the neural networks for $\log_{10} \mstar$. Input column codes are: m, magnitudes; c, pairwise colors; C, all colors.  The best-performing set of inputs for the noise-free case (the set with the minimum loss function) is highlighted in bold text. 
See section~\ref{sec:nnperf} for more details on column choices and performance metrics.}\label{tab:nnperf_mstar}
\end{center}
\end{table*}

\begin{figure}
	\includegraphics[width=0.45\textwidth]{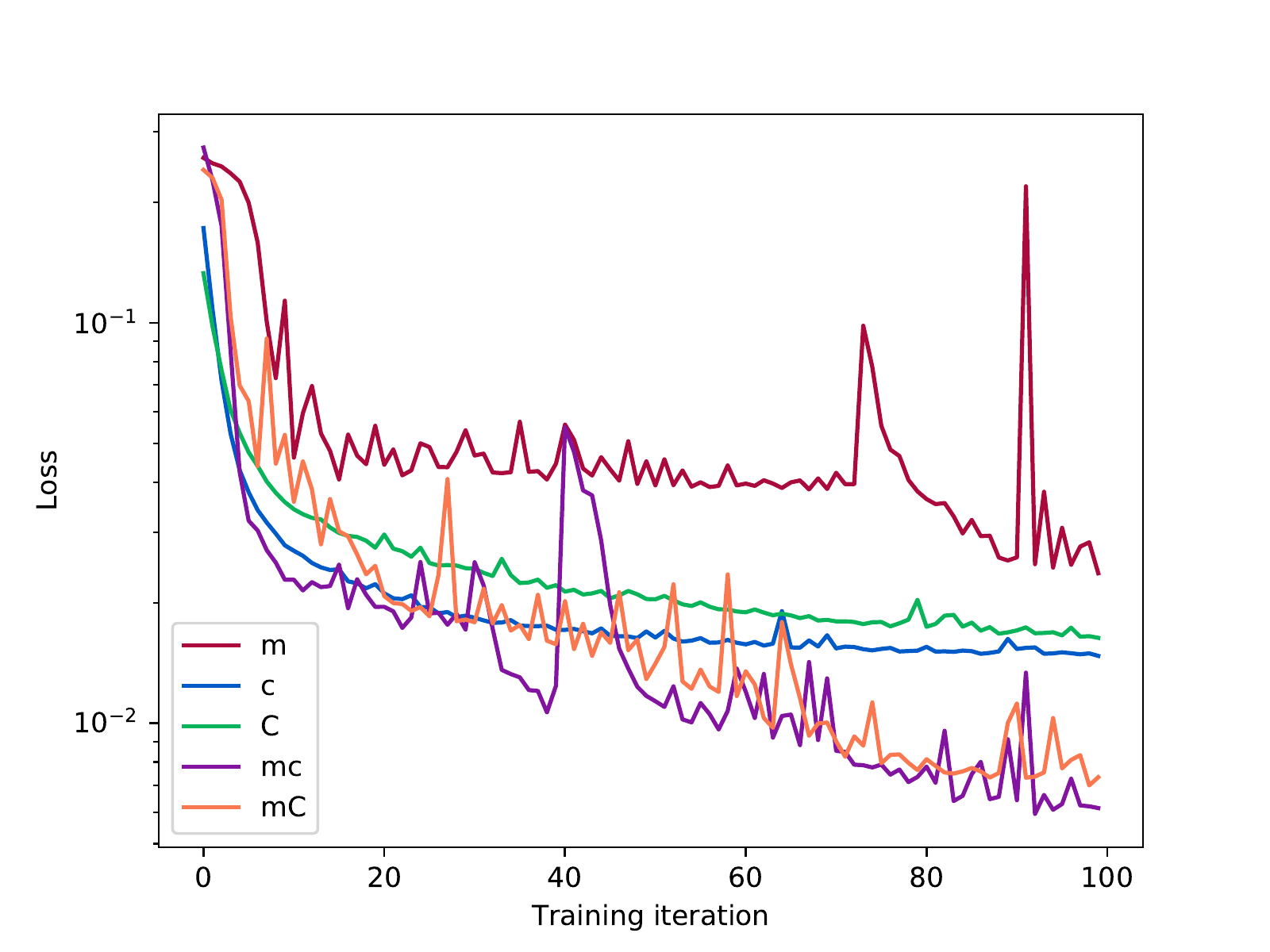}
	\caption{The loss function for the test set the first 100 training iterations (500 steps per iteration) in neural networks trained to reproduce stellar mass. Letter codes are different input column choices: m, magnitudes; c, pairwise colors; C, all colors.  Training the network is a numerical optimization problem and does not always proceed monotonically.}\label{fig:loss}
\end{figure}

Table~\ref{tab:nnperf_mstar} shows that we achieve the best performance when our input data includes galaxy magnitudes as well as the set of all colors (not just pairwise colors).  Again, while these combinations are possible for a neural network architecture to uncover, for our neural network architecture, we found that it was more efficient to provide all combinations at the beginning and let the network learn how best to include them, rather than letting those combinations be created by the network itself.

In Fig.~\ref{fig:loss}, we show the loss function plotted against the number of training steps for different sets of input columns.  The numerical noise of the algorithm is obvious from the non-monotonic behavior and sharp jumps in some of the lines. The magnitude-only or color-only networks, which did not perform as well for stellar mass, asymptote to relatively high values of the loss relatively early on; 100 steps in, however, the magnitude plus color networks are still improving.

\begin{figure*}
	\includegraphics[width=0.9\textwidth]{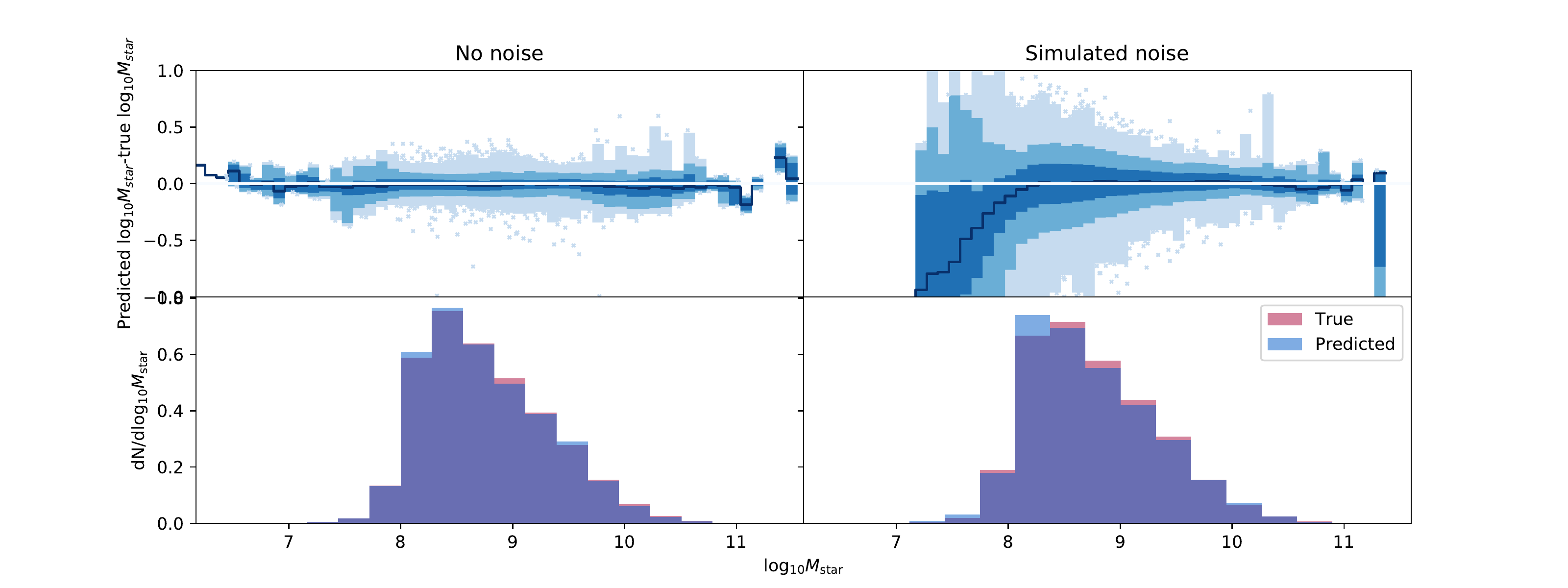}
	\caption{Neural network performance reproducing stellar mass on the validation set of the simulation galaxies. Left panels are noise-free, right panels have simulated observational noise \new{(the GOODS-S mock with uncertainties)}. Top panels: Prediction error relative to the true values as a function of predicted stellar mass. Performance is shown in bins of size $\Delta \log_{10} \mstar=0.1$, and the median and 1-, 2-, and 3$\sigma$ percentile contours are shown as the dark blue line and the three lighter blue regions, respectively.  Outliers are shown as light-blue crosses.  The white line is zero bias.  The bias is low in the intermediate range of stellar masses, where most of the training points are, but increases at low and high stellar mass.  Bottom panel: The histograms of predicted and actual stellar mass, which are similar in shape for the noise-free case, though the predicted stellar mass peaks at a slightly lower value than the truth.  The predicted distribution is significantly skewed in the noisy case relative to the expected distribution.}\label{fig:mstarperf_contour}
\end{figure*}

In the absence of noise, our performance in $\log_{10} \mstar$ is reasonable for individual galaxies, with overall biases a few percent or less and uncertainty of the order of 10\%, as shown in Table~\ref{tab:nnperf_mstar}.  Visual inspection confirms that the overall distribution is well reproduced, although the K-S test indicates that the distributions are statistically different.  The error as a function of galaxy parameter in Fig.~\ref{fig:mstarperf_contour} demonstrates that we can train the networks well where there is a high density of points, with increased errors towards the tails of the distribution.  

We now compare our results to stellar mass measurements of mock catalogs. \citet{2015ApJ...808..101M} performed a comprehensive study to estimate uncertainties and sources of bias in measuring physical parameters in galaxies.  A number of different tests were performed, based on different initial parameters and codes. Here we adopt their TEST-2A as our benchmark comparison. This is based on semianalytic models with a diversity of star formation histories and other parameters, and as with our semianalytic models, used measurements in CANDELS filter bands as their data set.  We note that TEST-2A used more CANDELS bands than our measurement here: 13 instead of 8, including U-band, K-band, and Spitzer infrared data in addition to the Hubble optical and near-infrared bands we use, and excluding the F184W (ACS) filter that we use.  Comparing to the distribution of biases and uncertainties returned by the individual methods in \citealt{2015ApJ...808..101M}, we find that we easily improve on the bias and have competitive uncertainty in our measurements.  However, we expect to have lower bias: different assumptions of initial mass functions, star formation rates, etc were found by \citealt{2015ApJ...808..101M} to be a source of systematic offsets between different codes, adding a constant bias to every galaxy, whereas we are implicitly assuming the same initial mass function and star formation rate since our samples are drawn from the same mock catalog. Our templates also have varying amounts of dust and varying metallicities, the lack of which has also been found to bias stellar mass estimates \citep{2013MNRAS.435...87M}.

We repeat the analysis with pseudo observational error, either matched to the full uncertainty distribution or matched to the uncertainty distribution of galaxies which have at least a $5\sigma$-detected flux measurement in each band. We show only the best-performing input column set from the noise-free case, after checking that this column choice still performs best when error is added.  This additional simulated observational noise causes an increase in bias and a large increase in uncertainty, visible in the right-hand plots of Fig.~\ref{fig:mstarperf_contour}.  However, the average bias of $-0.019$ dex is still smaller by an order of magnitude than any of the biases from~ \citet{2015ApJ...808..101M} (with the same caveat about correct implicit assumptions), and the uncertainty only a little worse than the maximum uncertainty from that comparison, 0.203 dex instead of 0.183 dex. \new{Because the simulation with full pseudo observational uncertainty is the closest to the real data and will be reused throughout the paper, we will call it the "GOODS-S mock with uncertainties", to distinguish it from the base mocks that do not include observational error or include a lesser amount of uncertainty.}

\subsubsection{Metallicity}\label{sec:metallicity}

\begin{table*}
\begin{center}
	\begin{tabular}{cccccccc}

 Noise & Input & Number &  & Average &  & 3$\sigma$ outlier\\
 level & columns & of steps & Loss$^{1/2}$ & bias & Uncertainty & rate \\
\tableline
None & m & 100000 & 0.092 & 0.002 & 0.092 & 0.013 \\
None & c & 100000 & 0.075 & 0.010 & 0.074 & 0.015 \\
\textbf{None} & \textbf{C} & $\mathbf{100000}$ & $\mathbf{0.064}$ & $\mathbf{-0.001}$ & $\mathbf{0.064}$ &  $\mathbf{0.017}$ \\
None & mc & 100000 & 0.070 & 0.007 & 0.069 & 0.014 \\
None & mC & 100000 & 0.065 & 0.017 & 0.062 & 0.0137 \\
		\tableline 
5$\sigma$ & C & 10000 & 0.149 & -0.008 & 0.149 & 0.011 \\
Full noise & C & 10000 & 0.204 & -0.035 & 0.201 & 0.015 \\

\end{tabular}
	\caption{Summary of the performance of the neural networks for $\log_{10} \zstar$. Input column codes are: m, magnitudes; c, pairwise colors; C, all colors. The best-performing set of inputs for the noise-free case (the set with the minimum loss function)  is highlighted in bold text. 
See section~\ref{sec:nnperf} for more details on column choices and performance metrics.}\label{tab:nnperf_zstar}
\end{center}
\end{table*}

\begin{figure*}
	\includegraphics[width=0.9\textwidth]{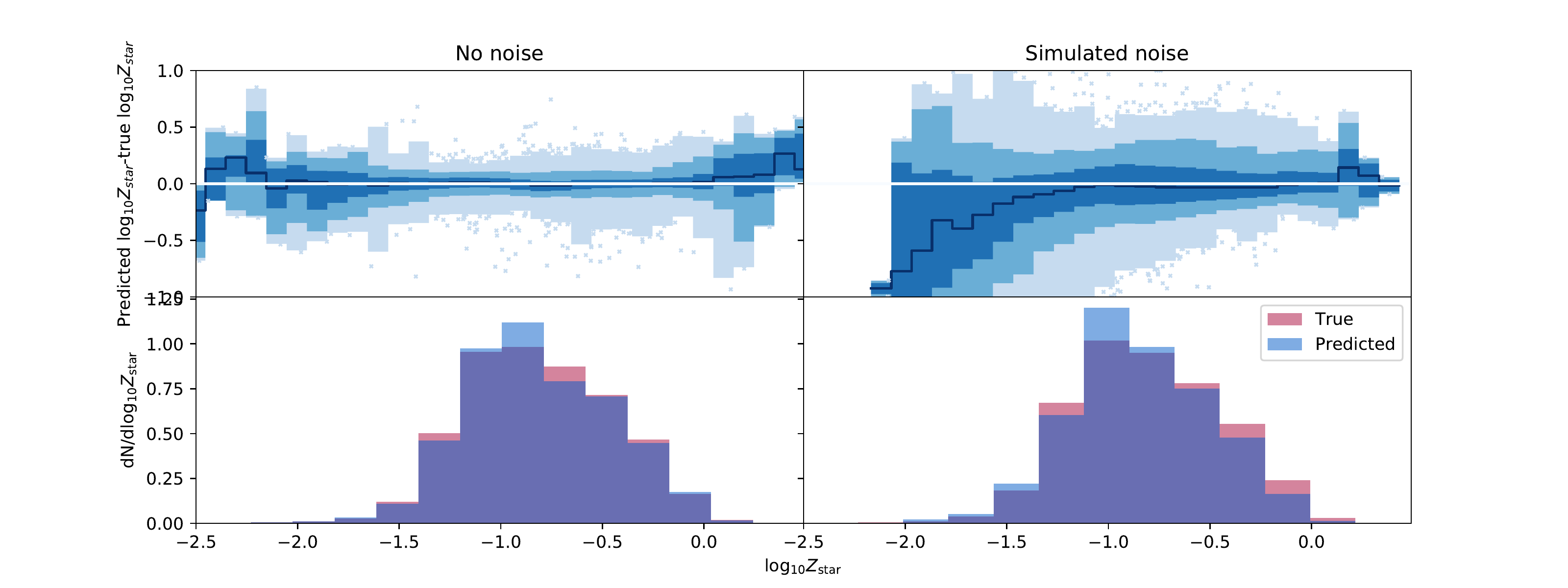}
	\caption{Neural network performance reproducing stellar metallicity on the validation set of the simulation galaxies. Left panels are noise-free, right panels have simulated observational noise \new{(the GOODS-S mock with uncertainties)}. Top panel: Prediction error relative to the true values as a function of predicted stellar metallicity. Performance is shown in bins of size $\Delta \log_{10} \zstar=0.1$, and the median and 1-, 2-, and 3$\sigma$ percentile contours are shown as the dark blue line and the three lighter blue regions, respectively.  Outliers are shown as light-blue crosses.  The white line is zero bias.  As with stellar mass, we perform well for metallicities in the middle of the metallicity range where the density of training points is high, but we see increased bias and, here, increased uncertainty for points with low or high metallicity.  Bottom panel: The histograms of predicted and actual stellar metallicity.  Here, the predicted metallicities have a narrower distribution than the true metallicities even in the noise-free case, although the amount of the difference is small--about 2 per cent in $\log_{10}{\zstar}$ space.}\label{fig:zstarperf_contour}
\end{figure*}

As with stellar mass, we find good performance from our metallicity-predicting neural networks, with small biases and order 10\% uncertainty, as shown in Table~\ref{tab:nnperf_zstar} and Fig.~\ref{fig:zstarperf_contour}.  The overall predicted distribution is somewhat more skewed from the original distribution than in the stellar mass case, though again we perform well where the density of training points is high.  As with the stellar mass, we perform worse when \new{we use the GOODS-S mock with uncertainties}, but not by a significant amount: an increase of bias by a factor of 2 and uncertainty by a factor of 3.  This uncertainty, of 0.2 dex in the full-noise case, is larger than the convolutional neural network machinery of \citet{WuBoada}, who obtained 0.08 dex uncertainty, albeit using substantially more data (3-color 128x128 pixel cutouts, not eight measured magnitudes) and brighter and lower-redshift galaxies (brighter than 25th magnitude).  We obtain a similar result to theirs in that low-metallicity galaxies have systematically high metallicity predictions in the presence of noise.

\begin{figure}
	\includegraphics[width=0.45\textwidth]{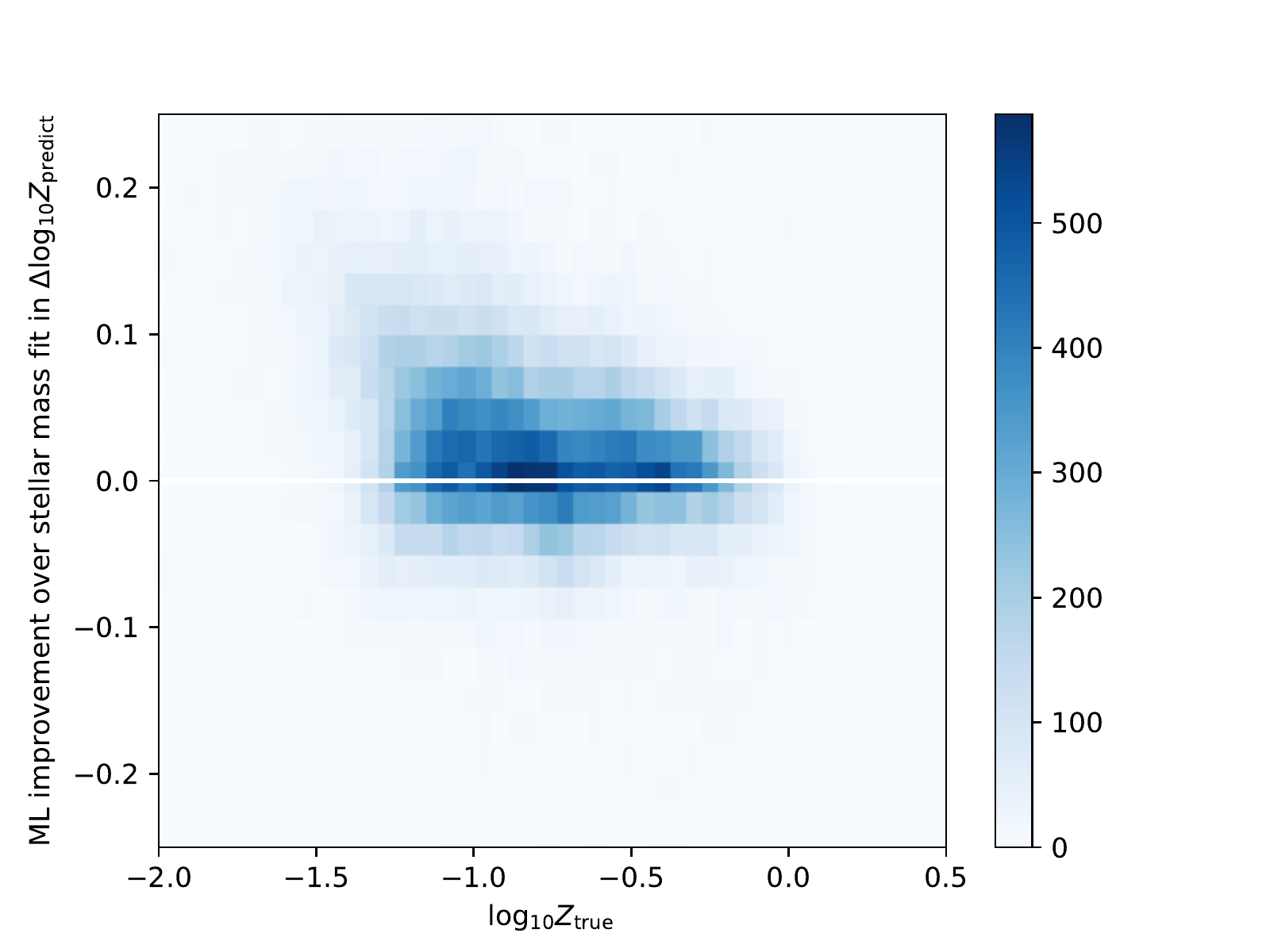}
	\caption{A 2-D distribution showing the improvement in metallicity estimation made by our neural network, relative to a simple model that uses the mean $\mstar-\zstar$ relation. Points above the white line at 0 have a neural network prediction that is closer to the true value of $\zstar$ than the simple model prediction.  The bulk of the points are above 0, indicating that the neural network is improving on the simple model, but the amplitude of the improvement indicates that the bulk of the correlation is being driven by the $\mstar-\zstar$ relation and not an independent measurement of metallicity. }\label{fig:mstar-zstar}
\end{figure}

This good performance on metallicity requires some discussion.  Metallicity is typically the hardest of the physical properties of galaxies measured by traditional methods.  The dominant component of our performance is driven by the relationship between metallicity and stellar mass.  There is a strong relationship between these quantities in data and in simulation~\citep[and references therein]{2014ApJ...795..123L}, and in practice the fact that we can fit stellar mass well means that most of our prediction of metallicity comes from the ability of the network to learn the $\mstar-\zstar$ connection and to predict $\mstar$ from the photometry, as can be seen by the similar shape of the bias-parameter curves in the noisy right-hand columns of in Fig.~\ref{fig:mstarperf_contour} and Fig.~\ref{fig:zstarperf_contour}.  That cannot be the full picture--metallicity performs best when we use colors alone, while stellar mass performs best when we use both magnitudes and colors--but it is a significant contributor to our good results.  To illustrate this, we fit a simple quadratic equation to the $\log_{10} \mstar-\log_{10} \zstar$ relationship in the simulations to predict a metallicity for each galaxy assuming no noise, and then we check how much improvement we get from the noise-free neural network when compared to this simple $\mstar$-based prediction. Fig.~\ref{fig:mstar-zstar} shows a 2-D histogram of this improvement, with points above 0 being an improvement on the polynomial fit, and points below 0 having an increased bias.  The bulk of our neural network predictions do improve on the simple model, by 1-2$\sigma$, but given that our data span more than 2 orders of magnitude, an improvement of 0.1-0.2 dex indicates that the simple stellar mass-metallicity relationship can explain a good fraction of our predicted metallicities.  The additional improvement is also suggested by the fact that the metallicity network performs better without magnitudes, while the stellar mass network performs better with them, meaning that the metallicity network must be learning different information than stellar mass alone.  Still, this effect suggests that, for any method, if metallicity is not well-constrained by the data, using a stellar mass-based prior with appropriate uncertainty may produce satisfactory results.

\subsubsection{Star formation rate}

\begin{table*}
\begin{center}
	\begin{tabular}{ccccccccc}

Noise &  Input & Number &  & Average &  & 3$\sigma$ outlier \\
level & columns & of steps & Loss$^{1/2}$ & bias & Uncertainty & rate \\
\tableline
None & m & 100000 & 0.297 & $-0.063$ & 0.291 & 0.016\\
None & c & 100000 & 0.178 & $-0.014$ & 0.178 & 0.015 \\
None & C & 100000 & 0.179 & $-0.004$ & 0.179 & 0.016 \\
None & mc & 100000 & 0.142 & 0.019 & 0.141 & 0.014 \\
\textbf{None} & \textbf{mC} & $\mathbf{100000}$ & $\mathbf{0.150}$ & $\mathbf{0.053}$ & $\mathbf{0.141}$ & $\mathbf{0.012}$ \\
\tableline		
5$\sigma$ & mC & 100000 & 0.213 & $-0.043$ & 0.227 & 0.018 \\
Full noise & mC & 100000 & 0.296 & $-0.088$ & 0.304 & 0.023 \\

\end{tabular}
	\caption{Summary of the performance of the neural networks for $\log_{10} \SFR$. Input column codes are: m, magnitudes; c, pairwise colors; C, all colors. The best-performing set of inputs for the noise-free case (the set with the minimum loss function) is highlighted in bold text. 
See section~\ref{sec:nnperf} for more details on column choices and performance metrics.}\label{tab:nnperf}
\end{center}
\end{table*}

\begin{figure*}
	\includegraphics[width=0.9\textwidth]{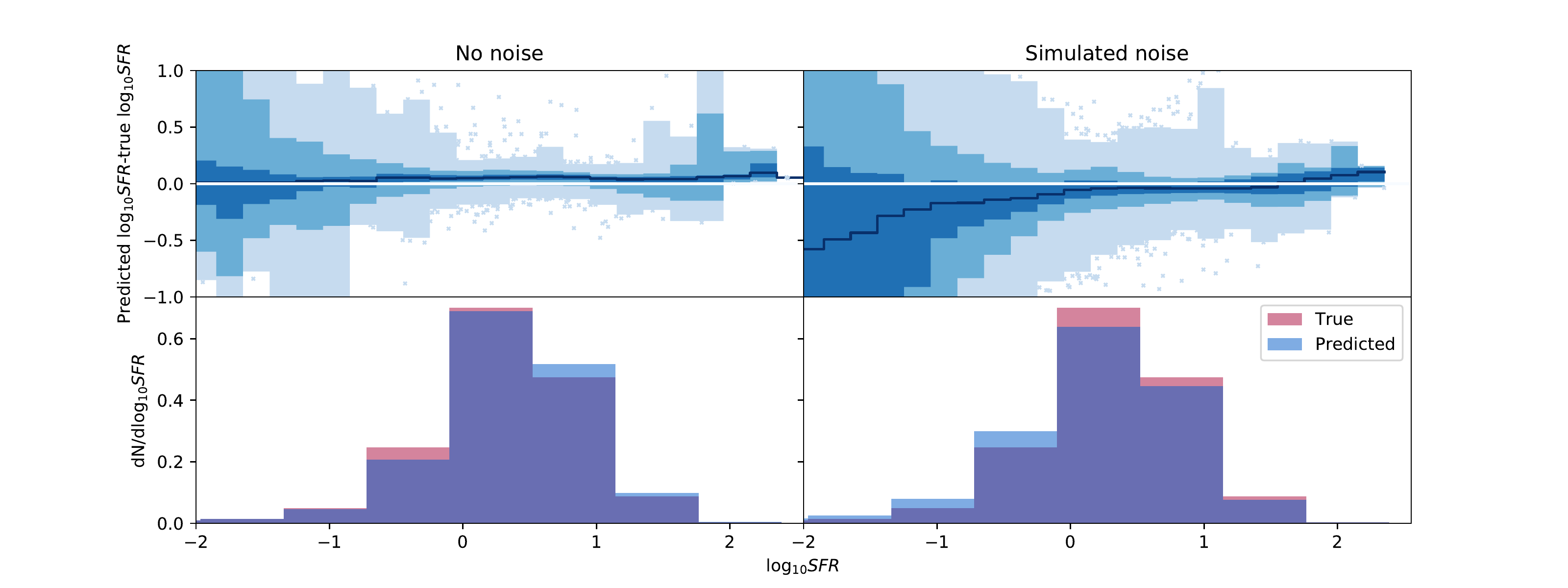}
	\caption{Neural network performance reproducing average star formation rate on the validation set of the simulation galaxies.  Left panels are noise-free, right panels have simulated observational noise \new{(the GOODS-S mock with uncertainties)}. Top panel: Prediction error relative to the true values as a function of predicted average star formation rate.  As in the above figures, the median and 1-, 2-, and 3$\sigma$ percentile contours are shown as the dark blue line and the three lighter blue regions, respectively.  Outliers are shown as light-blue crosses.  The white line is zero bias.  The large uncertainty in average star formation rate can be seen clearly, with an order of magnitude in uncertainty for the least star-forming galaxies.  A bias is visible even in the well-constrained regions near the peak of the star formation rate distribution.  Bottom panel: The histograms of predicted and actual star formation rate.  The histograms are not markedly different, although the difference is statistically significant given the large number of objects in each bin.}\label{fig:sfrperf_contour}
\end{figure*}

For the star formation rate, we have slightly higher bias and uncertainty.   The bias of $-0.053$ dex corresponds to a 12 percent error in linear space, visible in the histogram of Fig.~\ref{fig:sfrperf_contour} as the blue prediction histogram peaking slightly higher than the true distribution in red.  Much of the uncertainty is contributed by points with very low star formation rates. Fig.~\ref{fig:sfrperf_contour} shows the performance as a function of predicted star formation rate as well as a histogram of true and predicted values.  

This increased bias relative to other physical parameters is consistent with results from the literature: \citet{2019MNRAS.486.5104L}, performing a similar comparison to ours using SED fitting and the Horizon-AGN hydrodynamic simulations \citep{2017MNRAS.467.4739K}, find star formation rate biases of order 0.2 dex and uncertainties of order 0.2-0.6 dex, very consistent with our full-noise bias of $-0.09$ dex and uncertainty of $0.3$ dex.  We expect to do better than \citealt{2019MNRAS.486.5104L}, however, since one of the limitations of the SED fitting approach is that a (usually simplistic) star formation history must be assumed, while our training sample is instead drawn from a simulation which allows for more variation in star formation history.

\subsection{Redshift effects}

\begin{table*}
    \begin{center}
	\begin{tabular}{cccccccc}
		&  Input & Number &  & Average &  & &\\
		Property & columns & of steps & Loss$^{1/2}$ & bias & Uncertainty &$\Delta$ $|$bias$|$  & $\Delta$ uncertainty\\
		
		\tableline 
		$\mstar$ &mCz & 100000 & 0.046 & -0.010 & 0.045 & 0.001 & 0.011\\
		\tableline
		$\zstar$ &Cz & 100000 & 0.078 & 0.009 & 0.077 & $-0.008$ & $-0.013$ \\
		\tableline
		$SFR$ &mCz & 100000 & 0.136 & $-0.004$ & 0.136 & $-0.049$ & $0.005$  \\
	\end{tabular}
	\caption{Summary of the performance of the neural network for different predictions when redshift is included as an input column. Input column codes are as in Table~\ref{tab:nnperf}. We show only the best-performing input column set from the case without using redshifts to examine the improvement redshift provides. See section~\ref{sec:nnperf} for more details on column choices and performance metrics.  We note that the value of bias for star formation rate, $-0.004$, is a chance noise fluctuation down relative to the performance of the network, and the bias for a different subset of the data would likely be larger.}\label{tab:nnperf_redshift}
	\end{center}
\end{table*}

None of these networks used redshift as an input, but template-fitting methods generally require it \citep{2015ApJ...808..101M}.  Do our results improve if we add redshift as an input to the networks trained on the other galaxy properties?  We take the best-performing input column set for stellar mass, metallicity, and star formation rate, and add redshift as an input.  The results are summarized in Table~\ref{tab:nnperf_redshift}.  For stellar mass and star formation rate, the mean bias and the uncertainty both improve.  However, the improvement in the uncertainty is generally small, and the change in bias is less than the uncertainty for all three galaxy properties.  This is an encouraging result, as it means we can achieve good accuracy without needing to supplement our data with spectroscopic redshifts.

\section{Comparison of SED-fitting and neural networks on mock data}\label{sec:nn_sed_comp}

\new{We would like to use our trained neural networks on real data. However, real data lacks perfect information on the quantities we are measuring. For our comparison data set, then, we will measure our physical parameters of interest through template fitting.} 

\begin{figure*}
\begin{tabular}{cc}
         \includegraphics[width=0.45\textwidth]{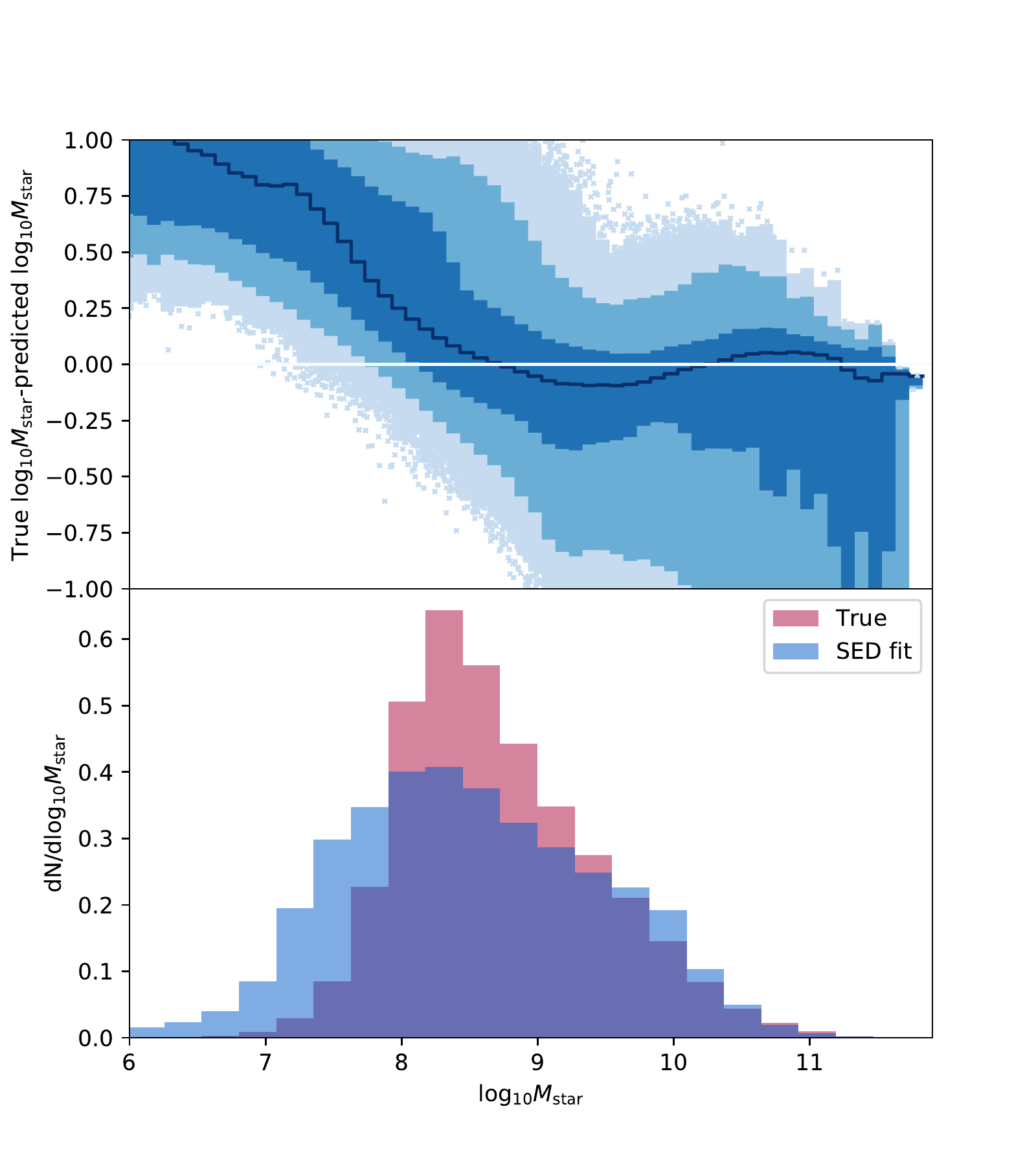} &
         \includegraphics[width=0.45\textwidth]{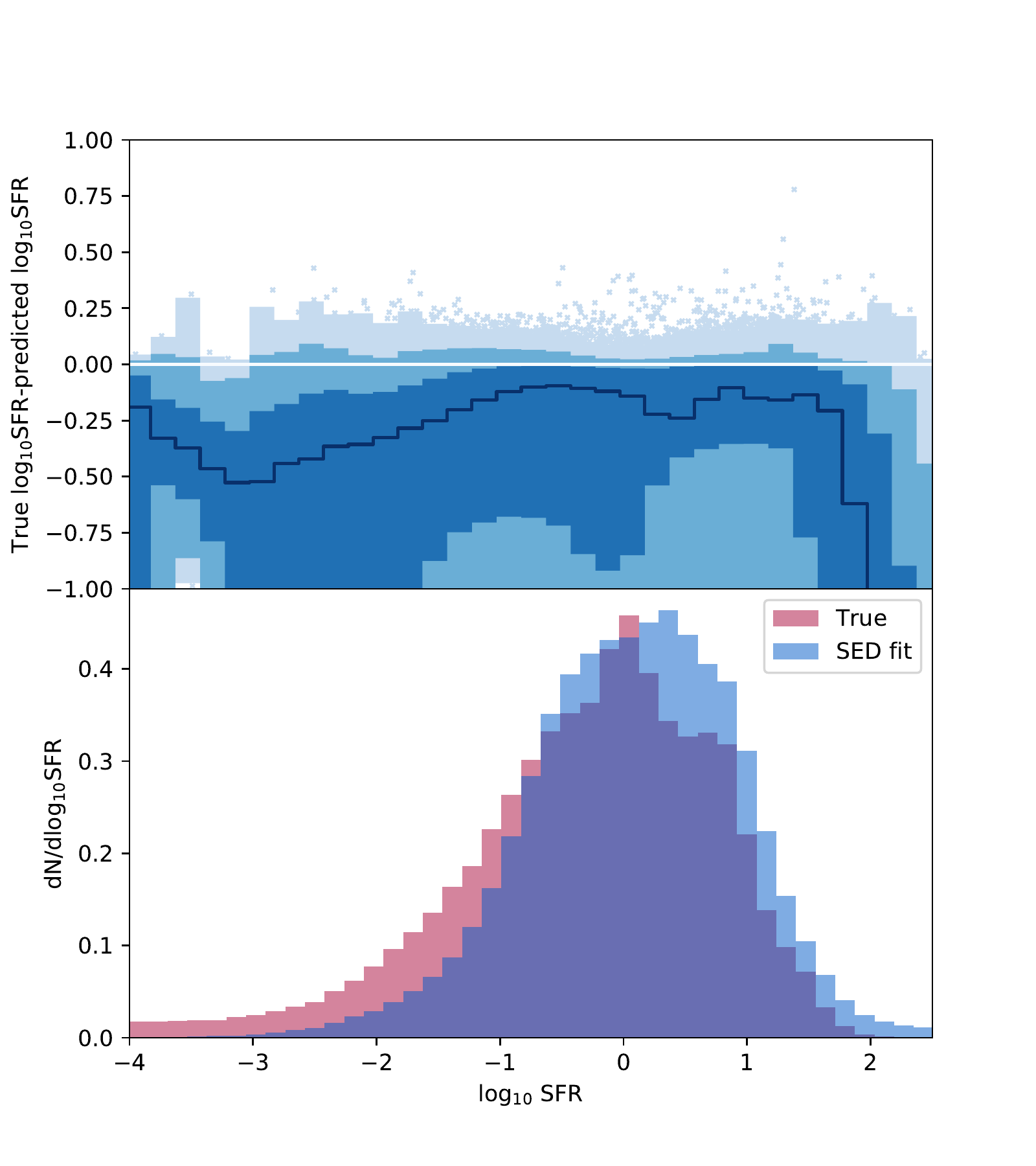}
\end{tabular}
	\caption{Summary statistics of performance of the SED fitting procedure on the GOODS-S mock with uncertainties.  Since we know the true underlying value of these parameters, any differences are the result of different parameterizations in associating the photometry with the mock catalogue or with the templates.  We see offsets at low stellar mass and globally for all star formation rate values; the differences are consistent with expectations for how large the effect of parameterization should be. (The sign of the star formation rate bias is different from \citet{2019MNRAS.486.5104L}, but we have a significantly different mass and redshift distribution, and the sign of the effect is different in different mass and redshift ranges.) We note that this figure plots the true value minus the predicted value, rather than the predicted value minus the true value as in other works in this paper, to facilitate comparison with Fig.~\ref{fig:dataperf_noise} where we plot the neural network result minus the SED fit, \new{since we do not have a known true value. If the neural network produces the same thing as the true result, in other words, Fig.~\ref{fig:dataperf_noise} will look identical to these plots}.}\label{fig:fitmock}
\end{figure*}

\begin{table}
\begin{center}
    \begin{tabular}{cccc}
    & Average & & 3$\sigma$ outlier \\
    Property & bias & Uncertainty & rate \\
    \tableline
	$\mstar$ & -0.193 & 0.424 & 0.004 \\
	$SFR$ & 0.490 & 1.06 & 0.031
    \end{tabular}
    \caption{\new{Summary of the performance of the SED-fitting procedure on the GOODS-S mock with uncertainties, shown in Fig.~\ref{fig:fitmock}. The bias in stellar mass comes largely from low stellar mass values, while the large scatter for stellar mass and the large bias and scatter for star formation rate are found across the entire range of values. Note that the bias is given as (predicted-true), like other tables in this paper, although the figure plots (true-predicted) for ease of comparison to later figures.}}\label{table:fitmock}
\end{center}
\end{table}

\begin{table*}
\begin{center}
	\begin{tabular}{ccccccc}
		
		&  Input & Number & Noise & Average &  & 3$\sigma$ outlier\\
		Property & columns & of steps & property & bias & Uncertainty & rate\\
		
		\tableline 
		$\mstar$ & mC & 10000 & No noise & 0.270 & 0.950 & 0.014 \\
		$\mstar$ & mC & 10000 & 5$\sigma$ cut & 0.208 & 0.581 & 0.029 \\
		$\mstar$ & mC & 10000 & Full noise & 0.149 & 0.559 & 0.023  \\
        \\
		$SFR$ & mC & 10000 & No noise & $-1.06$ & 2.14 & 0.019 \\
		$SFR$ & mC & 10000 & 5$\sigma$ cut & $-0.591$ & 1.43 & 0.029  \\
		$SFR$ & mC & 10000 & Full noise & $-0.605$ & 1.37 & 0.031 \\
		
	\end{tabular}
	\caption{Summary of the performance of the neural network trained on the GOODS-S mock with uncertainties and applied to CANDELS data for different predictions. Input column codes are as in Table~\ref{tab:nnperf}.}\label{tab:dataperf_noise}
\end{center}
\end{table*}

Here we quantify the difference expected in the analysis \new{due to the difference between the true values and the template-fitting results}. Different underlying parameterizations or parameter choices in the mock catalog simulations and the SED templates (e.g., star formation histories and dust attenuation laws) may lead not just to an increase in RMS uncertainty or a global offset but to biases in different regions of the parameter space.  To explore this effect, we perform the same SED template fitting procedure directly on the mock catalog for the GOODS-S region, doped with observational uncertainty as described previously, \new{which we continue to call} "the GOODS-S mock with uncertainties".  We show the results for stellar mass and star formation rate in Fig.~\ref{fig:fitmock}, with the result summarized in table form in Table~\ref{table:fitmock}.  We find a clear offset at low stellar mass and a global offset in the star formation rate.  The size is about as expected above $10^8 \mstar$, and somewhat larger for the less-frequent small stellar mass galaxies.  The offset in star formation rate is also as expected, $0.1-0.2$ dex. \new{Some of the large scatter is caused by the large number of faint galaxies in the semianalytic sample, where SED fitting does not perform as well; while the neural network performs better for these faint galaxies, on a more strongly magnitude-selected sample like we see in the CANDELS data, the performance of the two methods is similar. The scatter shown here, then, with the full complement of faint galaxies, will tend to underestimate the discrepancy between our neural network results and our SED-fitting results.}

\section{Performance on CANDELS data}\label{sec:mldata}

\new{Now,} We \new{turn to exploring} if networks trained on semianalytic models can be applied to observed galaxy data.  We use the CANDELS data from the GOODS-S field described in Section~\ref{sec:data:candels}, with the physical parameters measured through template fitting \new{as described in Section~\ref{sec:nn_sed_comp})}.  From \citet{2015ApJ...808..101M}, we know that different template-fitting methods can result in differences in the stellar masses of 0.2 dex. From \citet{2019MNRAS.486.5104L}, we expect that our star formation rate biases will also be of order $0.2$ dex with greater uncertainty. We may be able to improve on star formation rate in particular, since using a simulation has allowed us to use more complicated star formation histories than a typical SED fitting approach, and the simplicity of star formation histories is a major source of bias and uncertainty in SED-derived star formation rates. However, we are limited in the dust attenuation laws used in our simulation. Since this is also a major source of uncertainty in star formation rates \citep{2019MNRAS.486.5104L}, our ability to improve on star formation rate performance may be similarly limited. Both of these give an estimate of the accuracy we should aim for in our estimates of the stellar mass and star formation rate by our neural network technique.

\begin{figure*}
\begin{tabular}{cc}
         \includegraphics[width=0.45\textwidth]{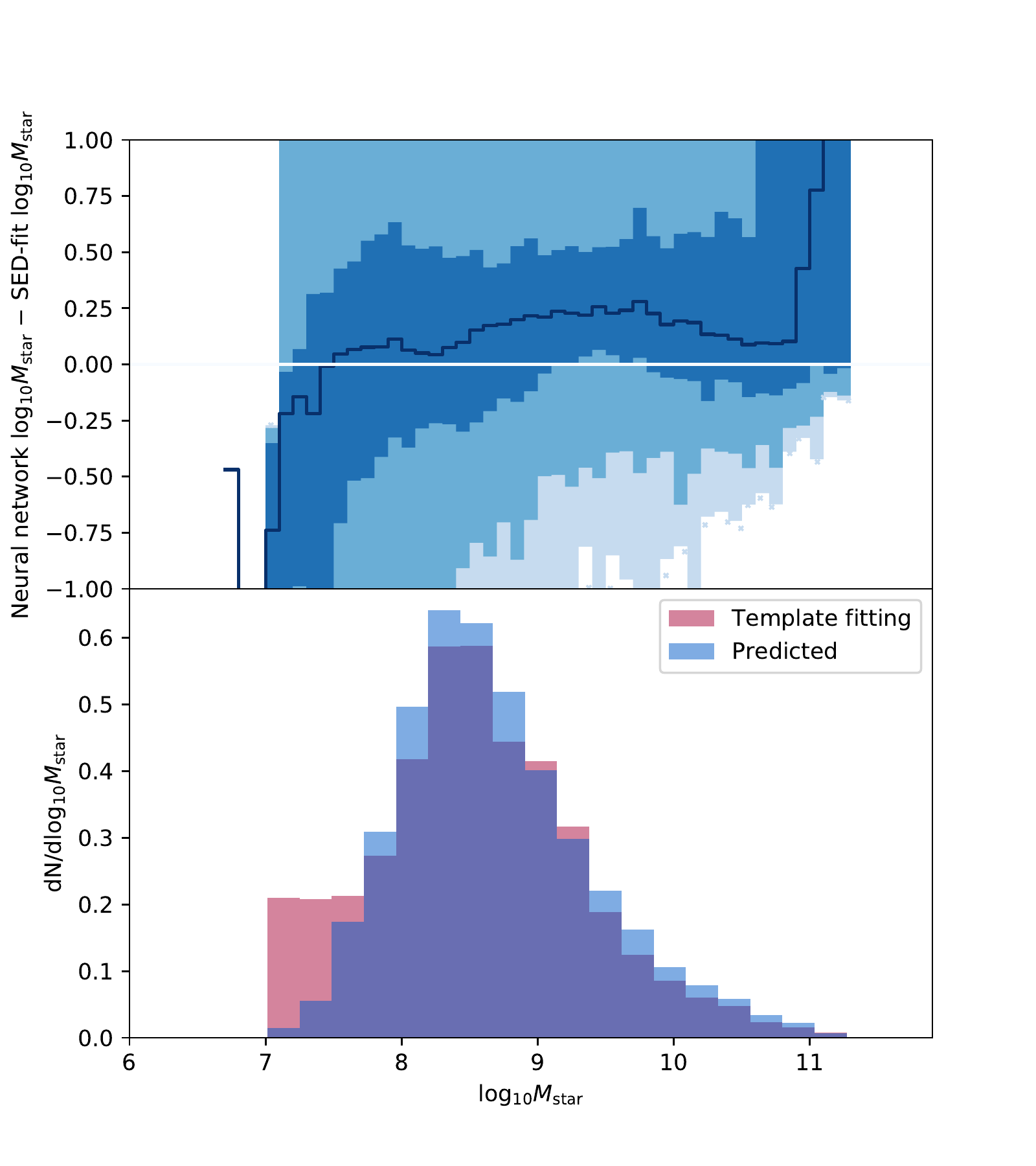} &
         \includegraphics[width=0.45\textwidth]{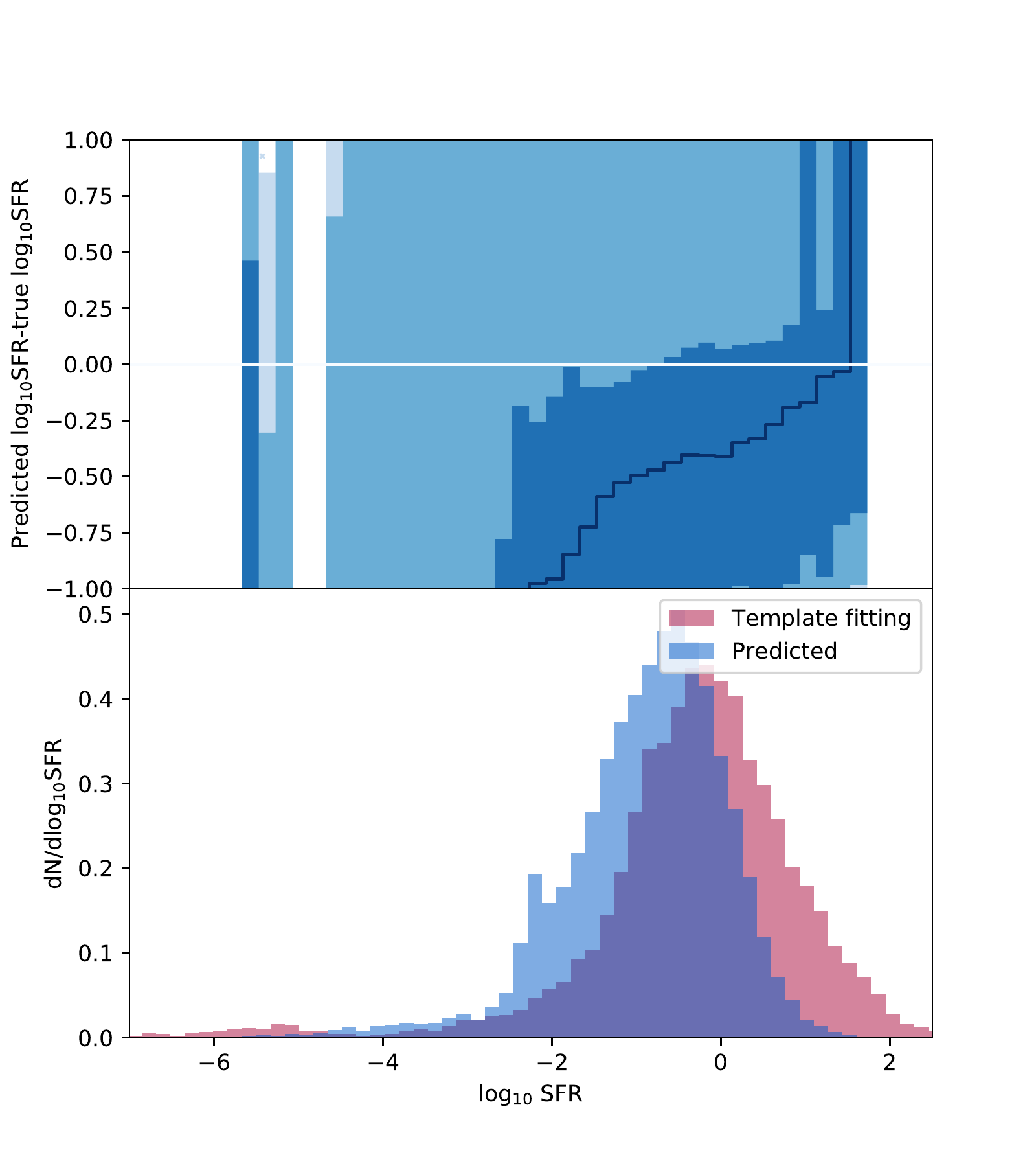}
\end{tabular}
	\caption{Summary statistics of performance of the neural network trained on the GOODS-S mock with uncertainties and applied to CANDELS data. In each subfigure, the top panel shows the prediction error relative to the SED-fit values as a function of the predicted value for the galaxy property, and the bottom panel shows a histogram of the distributions of the galaxy property in the SED fits and in our predictions. As before, in the top panel, the blue bars are the 1-, 2-, and 3$\sigma$ percentile contours, while the median is in dark blue and the white line is zero error.  The left-hand pair is stellar mass, and the right-hand pair is average star formation rate.  Because the distributions are not symmetric about the median, the median biases shown in this figure are not equal to the mean bias from Table~\ref{tab:dataperf_noise}.}\label{fig:dataperf_noise}
\end{figure*}

We show the performance on CANDELS data of the network trained on the GOODS-S mock with uncertainties in Table~\ref{tab:dataperf_noise} \new{and Fig.~\ref{fig:dataperf_noise}}. The neural networks trained on semianalytic data without observational noise, or with limited observational noise, do not reproduce the data well.  This is not surprising, however: the power of neural networks is that they reproduce high-dimensional nonlinear structures, so perturbing the data by some error has nonlinear effects on the prediction.  As we add noise, the performance on the data improves, even though the performance on the mock catalogs degrades, since the training mock data becomes more similar to the observational data.

The difference between the neural network predictions and the template-fitting predictions for the stellar mass are within our expectation of 0.2 dex for differences between template-fitting methods with different assumptions.  We note that the difference is of the opposite sign as would be expected from Fig.~\ref{fig:fitmock}, where we expect the predicted stellar mass to be larger than the (underestimated) SED-fit stellar mass at low masses; exact comparisons cannot be made since we plot as a function of predicted stellar mass, meaning the curves from Fig.~\ref{fig:fitmock} are actually shifted by some amount in the $x$-direction.  Still, even including this effect, our median stellar masses are within the expectation. However, the star formation rate numbers are significantly more discrepant.  Here, the difference is in the expected direction from Fig.~\ref{fig:fitmock}, although larger; a rough estimate is that about half the observed bias is the bias from the SED fitting, meaning that we are likely within expectations for $\log_{10} \SFR \gtrsim 0$, but still above that bias value for smaller star formation rates.  The large uncertainty in stellar mass likely derives from the uncertainty in the SED fits, which is large, up to $\pm 0.75$ dex, similar to the uncertainty in this comparison.  Again, the star formation rate values have greater scatter than expected from either the SED fitting or from the neural network training, indicating that the trouble likely lies in a difference in the mock catalog relative to the data. 

Interestingly, we sometimes obtain lower average bias or uncertainty when the method is applied to CANDELS data than we did on the simulations.  This is because the neural networks, like most methods, perform better on brighter galaxies than fainter galaxies \new{since brighter galaxies have lower relative uncertainty than fainter galaxies. Therefore, when we compare our results on simulations to our results on CANDELS data, we are conflating two effects: one, the fact that we are not as accurate or precise for a given individual galaxy in the CANDELS data as we are for a given galaxy of the same magnitude in the mock data, and two, the fact that we are measuring an average performance across a population and the CANDELS galaxies are on average ``easier'' since they are brighter. Using a population of brighter galaxies like the CANDELS data, in other words, will produce an apparently better result for the exact same method than using a population of fainter galaxies like the simulations. In our case, the size of the difference between the average apparent magnitude of the two populations---simulations and CANDELS data---is} 0.2-0.35 magnitudes depending on the filter. \new{This difference in average population improves our performance so much, just by being an easier problem to solve, that our average performance improves, even if our performance on similar galaxies worsens.}

The performance in $\log_{10} \mstar$ is promising, but the performance of star formation rate is less satisfactory. We consider here what could be causing this difference. The first possibility is differences between the physics of the simulation, the physics of the SED-fitting procedure, and the physics of the universe. Because both the simulation and the SED-fitting procedure have certain complexities the other method lacks, such as variation in star formation histories in the simulation and emission lines in the SED-fitting, and because we do not have truth values for the universe, distinguishing subtle differences would require more detailed investigation than we undertake here. However, we can investigate systematic differences between the SED fitting or the underlying CANDELS data and our training sample or training procedure.  We can first rule out most causes that fault the SED fitting rather than the neural network: we know the approximate size of this uncertainty from our tests above.

The largest difference between the mock catalog and the observational data is that the mock catalog contains no noise.  We dope the mock catalog with noise and improve our performance, but we are working with a pre-existing mock catalog that had already applied selection cuts to match the CANDELS data.  Since the process of applying noise and applying cuts do not commute, we may have induced a kind of Eddington bias, where we are preferentially missing the type of faint galaxies (and therefore preferentially low mass and low star formation rate galaxies) that scattered up into our observational sample.  The templates, being a parametric set intended to work as a kind of basis function space, do not suffer from this bias, and therefore would be more accurate if Eddington bias is considered.  It is also possible that, while we correctly reproduce color-space density in one and two dimensions, we have incorrectly reproduced some higher-dimensional manifold when generating the noisy mock catalog, and therefore that more of our data points lie off the manifold than might be expected.  As mentioned above, the major strength but also the major weakness of neural network solutions is that they can reproduce highly nonlinear manifolds. When the data are a highly nonlinear manifold, this is good; however, even a small movement off the expected manifold can cause an extremely poor extrapolation.  The templates, being constrained, would not suffer this effect.  We made some measurements of average distance to the nearest training sample point for both the test set and the observational data, and did not see any signs of increased distance, but distance measures can be hard to interpret in high-dimensional space so this is still a possibility.  In the future, if an approach like this--ML algorithms trained on simulated data--is used to measure galaxy properties, a mock catalogue tuned not only to the physical parameters of interest but to the noise properties of the observational catalog will be needed to ensure accuracy. This is consistent with the findings of \citet{2019MNRAS.489.4817D}, who spend significant effort reproducing the noise properties of their observational data.

\section{Summary and Discussion}\label{sec:summary}

We train neural networks on semianalytic models to predict from photometric data three galaxy properties of scientific interest: stellar mass, stellar metallicity, and average star formation rate.  In the absence of noise--the best-case scenario--we achieve excellent accuracy and precision on all properties, indicating that the mapping from galaxy photometry to galaxy physical properties has low enough noise and few degeneracies.  Injecting artificial photometric noise degrades the performance on a reserved sample of semianalytic galaxies, but allows the algorithm to perform better on real data from CANDELS, which contains noise.

Our accuracy and precision show that semianalytic galaxies can be used to train neural networks that can produce photometric stellar masses, metallicities, and star formation rates, and that these galaxy properties are accessible targets for machine learning.  However, our performance on noisy simulated data is not yet fully competitive with mature template-fitting and machine-learning methods from the literature.  There are several likely reasons for this discrepancy.  Because our model includes many more free parameters, we may be trading off precision (a narrower set of outputs generated by a template) with accuracy: our mean bias is lower for stellar mass, but our uncertainty is higher.  Also, as mentioned in Section~\ref{sec:metallicity}, we have used a wide range of fluxes and redshifts, a calibration set that may be more ambitious than some of the data sets we are comparing to.  Still, as a first-step implementation of neural networks on these data, we have achieved good accuracy in predicting all galaxy properties.

\new{If the offset and scatter between neural networks and SED-fitting on simulation data is similar in real CANDELS data, then} the performance on real CANDELS data is not as accurate or precise as comparison data sets from the literature~\citep{2015ApJ...808..101M}.  Some of that is driven by our decreased accuracy in the presence of noise even for the semianalytic galaxies, although the fact that the galaxies in CANDELS are brighter, on average, than our semianalytic galaxies means that in some cases we do better on the average CANDELS galaxy than we do on the average semianalytic galaxy.  Also, we note that the photometry in the semianalytic catalog was designed to match the CANDELS observed distributions, but since the CANDELS data are noisy and the semianalytic galaxies are not, we expect some mismatch between the locations of the underlying noise-free manifolds. Additionally, we do not expect perfect agreement, since the template-fitting results rely on other parameters (such as redshift and metallicity) that may differ from the values in our semianalytic catalogue; these differences in assumptions explain some of our observed discrepancies. \new{Of course, some further underlying differences between the simulation and reality may make our comparison between the two methods less accurate. Assuming the measured relationship holds in CANDELS, then, and}  accounting for the measured bias and uncertainty between mock catalog values and the results from performing our SED fitting procedure on the mock catalogs, we have nearly competitive accuracy (in the sense of low bias) and precision (low scatter) in stellar mass, and competitive precision but larger-than-expected inaccuracy for star formation rate. For future research, we recommend that a study of this nature should use a mock catalog created for the purpose with photometric noise already included when generating the catalog, as this is the most likely cause of differences between our mock data and the CANDELS data.

Interestingly, our results show that adding redshift information to our training sample results in either minor improvement or in degradation of accuracy and precision, indicating that the networks are capable of learning the relevant mapping between color, redshift, and other galaxy properties without explicit redshift information.  This indicates that competitive accuracy and precision can be achieved even for galaxies that lack spectroscopic redshifts, greatly increasing the available sample sizes for future studies.

We have been able to obtain good accuracy and precision for stellar metallicities, which is the most difficult to measure of the galaxy properties.  This result is driven by a strong relationship between metallicity and stellar mass.  We improve on the accuracy obtained by simply relying on a stellar mass--metallicity relation, but the improvement is order 0.1 dex, compared to the more than two orders of magnitude range in the parameter space, indicating that the predictive accuracy is dominated by the stellar mass--metallicity relation.  This suggests the use of a strong stellar mass-metallicity prior when trying to obtain stellar metallicities from low-resolution data.

Future work will be needed to develop the machine learning architecture to a higher level of complexity and precision in order to be competitive in accuracy with currently-existing methods.  However, our work shows that simple machine learning methods that do not know about the physics of galaxy formation and evolution can in principle reproduce galaxy property measurements with high accuracy, as long as the training set and the data of interest are consistent with one another and the machine learning is carefully trained to handle degeneracies and noise inherent in the data.  This allows for not only the use of semianalytic models as a training set for galaxy property measurement, but also an increase of computational speed of template-fitting efforts--for example, if template-fitted stellar masses are available for a representative spectroscopic subset of a large photometric survey, then machine learning is a computationally efficient way to extend those template-fitting results to the larger photometric data set.  As we have shown here, machine learning methods can be trained on complicated relationships between galaxy properties and photometry measurements drawn from a small number of filters, indicating that scientifically interesting galaxy properties can be measured with reasonable computational time on future datasets from large-scale photometric surveys.
                                    
\acknowledgements

Support for this work was provided by the University of California Riverside Office of Research and Economic Development through the FIELDS NASA-MIRO program.  A portion of this research was carried out at the Jet Propulsion Laboratory, California Institute of Technology, under a contract with the National Aeronautics and Space Administration. We would like to acknowledge Dritan Kodra for the updated CANDELS photometric redshift catalog.  
                                             
\bibliographystyle{mnras}
\bibliography{ml_sam}	
\end{document}